\documentclass[11pt,hyper]{JHEP3}
\usepackage{amssymb,amsmath,amsfonts,cite}
\usepackage{epsfig}
\usepackage{graphicx}
\usepackage{dcolumn}
\usepackage{bm}
\usepackage{amsthm}
\usepackage{amsmath}
\usepackage{setspace}
\usepackage[normalem]{ulem}
\usepackage{relsize}
\usepackage{cancel}

\allowdisplaybreaks[1]

\usepackage{amsthm}

\newcommand{\del}{\partial}
\newcommand{\ddsimple}[2]{\frac{\del #1}{\del #2}} 

\title{Superfluid Kubo Formulas from Partition Function}

\author{Shira Chapman, Carlos Hoyos, Yaron Oz\\
Raymond and Beverly Sackler School of Physics and Astronomy, Tel-Aviv University, Tel-Aviv 69978, Israel\\
{\it E-mail:} \email{shirator@post.tau.ac.il, choyos@post.tau.ac.il, yaronoz@post.tau.ac.il}
}

\abstract{
Linear response theory relates hydrodynamic transport coefficients to equilibrium retarded correlation functions of the
stress-energy tensor and global symmetry currents in terms of  Kubo formulas.
Some of these transport coefficients are non-dissipative and affect the fluid dynamics at equilibrium. We present an algebraic framework  for deriving Kubo formulas for such thermal transport coefficients by using the equilibrium partition function.
We use the framework to derive Kubo formulas for all such transport coefficients of superfluids, as well as
to rederive Kubo formulas for various normal fluid systems.
}


\begin{document}





\section{Introduction and Outlook}

Hydrodynamics is the long wavelength effective description of a dynamical system at local  thermal equilibrium.
The fluid dynamics is governed by the conservation laws of the stress-energy tensor and charge currents, whose dependence on the thermal parameters, such as  the fluid velocity, temperature and chemical potentials,  is given by constitutive relations. These, supplemented by an equation of state define the hydrodynamics completely.

It has been recently demonstrated  in  \cite{Banerjee:2012iz,Jensen:2012jh}, that the non-dissipative properties of hydrodynamic systems are captured by the
equilibrium partition function on curved stationary backgrounds.
The most general gauge and diffeomorphism  invariant equilibrium partition function on such backgrounds consists of thermal functions, i.e.
functions of the temperature and the chemical potentials.
The hydrodynamic transport coefficients can be expressed in terms of the thermal functions. This yields relations among the transport coefficients, since there are in general more transport coefficients than thermal functions.
These relations between hydrodynamic transport coefficients coincide with the equality type constraints on the transport coefficients
that are obtained by imposing the local second law of thermodynamics.

Linear response theory relates hydrodynamic transport coefficients to retarded correlation functions of the stress-energy tensor and charge currents
of the microscopic theory by Kubo formulas. Thus, the Kubo formulas provide means to calculate the properties of field theories
in their  hydrodynamic regime.
A way to derive these Kubo formulas is to consider the hydrodynamic stress-energy tensor and charge currents on an external gauge and gravity background
and differentiate with respect to the metric and gauge fields perturbations (for recent relevant works see e.g. \cite{Moore:2010bu,Moore:2012tc}).
This typically requires to solve the hydrodynamic equations for the various fields (velocity, temperature, chemical potentials etc.) in terms of the background metric and gauge fields, and substitute the solution into the constitutive relations for the stress-energy tensor and the charge currents.

As an alternative to this differential method, we propose
in this work a new algebraic framework for deriving Kubo formulas for the thermal functions and the transport coefficients, by using the
equilibrium partition function on stationary gauge and gravity
backgrounds.
The partition function encodes the stress-energy tensor and the charge currents and their dependence on the metric and the gauge fields,
which can be used in the linear response theory in order to derive the Kubo formulas.

A study of hydrodynamic transport coefficients in parity non-preserving superfluids using the local version of the second law of thermodynamics was performed in \cite{Bhattacharya:2011tra} to first dissipative order, and generalized  in the parity odd sector for an arbitrary number of unbroken charges in \cite{Neiman:2011mj}.

In \cite{Bhattacharyya:2012xi} the partition function analysis was carried out for relativistic superfluids\footnote{
A superfluid is a fluid with a spontaneously broken global symmetry. The
low energy degrees of freedom include also  the  massless Goldstone mode.
The hydrodynamics of a superfluid consists of two motions: the motion of the normal part of the fluid, and the motion of the superfluid part. The superfluid velocity lies in the direction of the Goldstone phase gradient.} with one important difference in the formalism. Instead of using the equilibrium partition function, the authors of \cite{Bhattacharyya:2012xi} used the local effective action for the massless Goldstone field. The reason for using the effective action rather than the partition function itself in the analysis of superfluid transport coefficients is that the equilibrium partition function is not a local functional of the external fields, while the effective field theory for the Goldstone mode is local.\footnote{The partition function can be obtained from the local effective action by integrating out the Goldstone boson. In a theory in which the Goldstone dynamics is effectively classical, e.g. at large $N$, all the correlation functions can be computed using the solution to the equation of motion for the Goldstone boson in terms of the background fields. Like \cite{Bhattacharyya:2012xi} we will be working in this limit.}

We use the same framework and the results of \cite{Bhattacharyya:2012xi,Banerjee:2012iz}  to derive Kubo formulas for thermal transport coefficients of superfluids, as well as
to rederive Kubo formulas for various normal fluid systems.
Since Kubo formulas are eventually evaluated on a flat background with no external gauge fields, at the final stage of our analysis the solution for the gradient of the Goldstone phase will no longer be non-local, but rather a constant independent thermal parameter, such as the temperature $T$ and chemical potential $\mu$. We will denote it by $\xi^\mu$, and its transverse part by $\zeta^i$. To leading order in derivatives the effective action in the presence of a background gauge potential $A^i$ and metric $g^{0i}=-a^i$, and in the absence of redshift ($g_{00}=-1$) takes the form:
\begin{equation}
\begin{split}
S &\ = S_0 + S_1^{even}+S_1^{odd}+S^{anom}\ ,\\
S_0&\ = \int d^3 x\frac{1}{ T} P(T, \mu , \hat{\zeta}^2)\ ,\\
S_1^{even}&\  = \int d^3 x f \left[c_1 (\hat\zeta \cdot \del)  T
+ c_2 (\hat\zeta \cdot \del) \mu +c_3(\hat\zeta \cdot \del) \hat \zeta^2\right]\ ,\\
S_1^{odd} & = \int  d^3x \,  \hat\zeta \cdot (g_1 \, \del \times A +T g_2\, \del \times a) \ ,
\\
S^{anom} & = C  \int d^3 x \, A\cdot \left( \frac{\mu}{3T}  \partial \times A+\frac{\mu^2}{6T} \partial \times a \right),
\end{split}
\end{equation}
where $\hat\zeta$ is the Goldstone field (for the redshifted version see subsection \ref{subsec:supfl:prelimII}). All the integrals are carried over the three dimensional volume element. All the vectors are oriented in the spatial directions and are contracted using the transverse part of the metric.
$C$ is the anomaly coefficient, $P$ is the thermodynamic pressure function and $f=-2(\del P / \del \hat \zeta^2)$.
Using our method we derive  the Kubo formulas for the three non-dissipative parity even thermal functions $c_i, ~i=1,2,3$
and the two parity odd thermal functions $g_1, g_2$.\footnote{Closely related to $\sigma_8$ and $\sigma_{10}$ of \cite{Bhattacharya:2011tra} or to
$\alpha_{ab}$ and $\beta_a$ of \cite{Neiman:2011mj}.} These functions can be used to express all the superfluid non-dissipative transport coefficients (see subsection \ref{subsec:supfl:transports} for details).
They enter in the constitutive relations of the current as ($\nu=\mu/T$):
\begin{equation}\label{ji}
\begin{split}
J^{\mu} & = q u^\mu -f \xi^\mu +
\frac{T^2sf}{\epsilon+P}  P^{\mu\nu} \left[ c_1 \del_\nu T+c_2 \del_\nu \nu +c_3 \del_\nu \zeta^2 \right]
\\ &+ B^\mu\left(C \mu + 2Tg_1\right)
+ \omega^\mu\left(\vphantom{\frac{n^a}{\epsilon+P}} C\mu^2 + 4 g_1 \mu T - 2 g_2 T^2 \right)+\cdots ,
\end{split}
\end{equation}
where $q,s,\epsilon$ are the charge, entropy and energy densities respectively, $u^\mu$ is the fluid velocity, $P^{\mu\nu}$ is the transverse projector, $\omega^\mu$ is the vorticity and $B^\mu$ is the magnetic field.
The terms proportional to the $c_i$ coefficients in the first line of \eqref{ji} look similar to conductivities, but, as opposed to conductivities in a normal fluid, they appear at thermal equilibrium in the constitutive relations of the current derived from the effective action. However, they are canceled by the first order corrections to the superfluid velocity once the equation of motion for the Goldstone field has been solved.

The Kubo formulas that we derive in the parity even sector make use of correlation functions of the Goldstone phase gradient and another (composite) operator.\footnote{Instead of the stress-tensor/charge-current correlators used everywhere else in this paper.} The reason for this slightly different approach in the parity even sector is due to the influence of the non-local terms mentioned earlier.
We get the following Kubo formulas in the parity even sector:
\begin{equation}
\begin{split}
\label{intro:even_kubo}
& c_1  = - \frac{1}{T^2} \lim_{\vec k ,\omega \rightarrow 0}  \langle \phi(\vec k) \ T^{00}(-\vec k) \rangle_{\vec\zeta_0\perp\vec k}
\\
&  c_2 = - \lim_{\vec k ,\omega \rightarrow 0}  \langle \phi(\vec k) \ J^{0}(-\vec k) \rangle_{\vec\zeta_0\perp\vec k}
\\
&  c_3 = \frac{i}{2T\zeta_0^y} \lim_{\vec{k}\rightarrow 0 }\frac{\del}{\del k_x}\langle \del_x\phi (k_x) \ J^{y} (-k_x);
\vec\zeta \parallel \hat y \rangle\ ,
\\ &
 f \equiv- 2 \frac{\del P}{\del \zeta^2} =- \lim_{\vec{k},w \rightarrow 0} G^{x,x}_{\perp} (\vec k_z,-\vec k_z; \vec\zeta \parallel \hat y)  \ ,
\end{split}
\end{equation}
where every correlation function has to be calculated at $\vec \zeta_0\perp \vec k$ (the zero momentum limit should be taken at the last step).
In the parity odd sector we have:
\begin{equation}
\begin{split}\label{intro:odd_kubo}
\notag &
    g_1 = - \lim_{k_n\rightarrow 0} \sum_{ij}  \frac{i}{4Tk_n}\epsilon_{ijn}   \widetilde G^{i,j}_\parallel(k_n,-k_n) \biggr|_{\omega=0} - \frac{C}{2} \left( \frac{\mu}{T} \right)
    \ ,
\\
&
    g_2 = \lim_{k_n\rightarrow 0} \sum_{ij}  \frac{i}{2T^2k_n}\epsilon_{ijn}
      \left[ \widetilde G^{i,0j}_\parallel(k_n,-k_n)  -\mu \widetilde G^{i,j}_\parallel(k_n,-k_n) \right]\biggr|_{\omega=0}
    -\frac{C}{2}\left(\frac{\mu}{T}\right)^2
    \ .
    \end{split}
\end{equation}
$G(k_n,-k_n)$ is the correlator of stress-tensors and currents according to its superscripts with external momentum $k$ in the $n$-th direction only (the exact definition is given in equation \eqref{anom:Green_def}), the tilde stands for a correlator obtained from a variation of the covariant current and \emph{wherever the subscripts $\parallel$/$\perp$ appear, the spatial momentum is taken to be perpendicular/parallel to the direction of the superfluid phase gradient $\zeta_i$}.

The formula for $g_1$ found above seems to reinforce the suggestion of \cite{Neiman:2011mj} that $g_1$ ($\alpha_{ab}$ in \cite{Neiman:2011mj}) may be related to a $JJT$ type anomaly. We however cannot prove directly that it vanishes. This will be explained in detail in the discussion.
Finally, we note that the same thermal functions can be often extracted from different components of the stress-tensor or currents.
The correlators obtained should be consistent with one another, therefore we get identities between different retarded correlators of
the stress-energy tensor and charge currents. We present examples throughout the text.

This paper is organized as follows. In section \S\ref{sec:anom}, we present our method, and implement it for a charged anomalous fluid in 3+1 dimensions at first order in the derivative expansion.
In section \S\ref{sec:supfl} we derive Kubo formulas for superfluid transport coefficients.
In the discussion we comment on the interpretation of the Kubo formulas for the parity odd transport coefficients in superfluids.
In addition to the material presented in the main text, in the last two appendices we consider all the other cases of \cite{Banerjee:2012iz} and derive the relevant Kubo formulas.

\section{Anomalous Charged Fluid in 3 + 1 Dimensions} \label{sec:anom}
In this section we study 3 + 1 dimensional charged fluid dynamics up to first order in the derivative expansion. We take into account the effect of quantum anomalies.
We will derive Kubo formulas for the hydrodynamic transport coefficients of such a fluid using the most general equilibrium partition function. We start with some preliminaries (see \cite{Banerjee:2012iz} for a detailed discussion).

\subsection{Preliminaries}
We will be working with the most general \emph{stationary} metric and gauge-connection background:
\begin{gather}
\begin{split}\label{anom:eqMetric}
ds^2 = -e^{2\sigma(\vec{x})} \left(dt+a_i(\vec{x})dx^i\right)^2+g_{ij}(\vec{x})dx^i dx^j,
\end{split}\\
\begin{split}\label{anom:eqGauge}
\mathcal{A}=\mathcal{A}_0(\vec{x})dx^0+\mathcal{A}_i(\vec{x})dx^i,
\end{split}
\end{gather}
in the notations of \cite{Banerjee:2012iz}.

The most general (CPT invariant) equilibrium partition function for such a system is:
\begin{gather}\label{anom:partition}
\begin{split}
& \ln Z = W^0+W^1_{inv} + W^1_{anom} \\
& W^0 = \int d^3 x \sqrt{g_3} \frac{e^\sigma}{T_0} P(T_0 e^{-\sigma}, e^{-\sigma} A_0), \\
& W^1_{inv} = \frac{T_0 C_2}{2} \int Ada,
\quad W^1_{anom} = \frac{C}{2} \int \left(\frac{A_0}{3T_0} AdA + \frac{A_0^2}{6T_0} Ada \right),
\end{split}
\end{gather}
where,
\begin{align}
\begin{split}\label{anom:KK_inv_gauge}
& A_0 \equiv \mathcal{A}_0+\mu_0\ , \\
& A_i \equiv \mathcal{A}_i-A_0 a_i\ ,
\end{split}
\end{align}
$\mathcal{A}$ is the gauge field, $\mu_0$ and $T_0$ are the equilibrium chemical potential and temperature used to evaluate the partition function.
$P(T,\mu)$ is the thermal pressure function, and
\begin{align}
\begin{split}
\frac{1}{2} \int XdY \equiv \int d^3x \sqrt{g_3} \epsilon^{ijk} X_i \del_j Y_k {}\ .
\end{split}
\end{align}
Since we are working on a stationary background, the partition function can be written as a three dimensional local integral.
The local values of the temperature and chemical potential are $T(\vec{x}) \equiv T_0 e^{-\sigma}$, $\mu(\vec{x}) \equiv A_0 e^{-\sigma}$, respectively. $C$ ,$C_2$ are constants. $C$ is the anomaly coefficient of the triangle diagram of three currents. It has been argued that $C_2$ is related to mixed gauge-gravitational anomaly \cite{Jensen:2012kj}.

Using the equilibrium partition function one derives the equilibrium stress-energy tensor and charge current:
\begin{align}
\begin{split}\label{anom:TJ_orig}
& T_{\mu\nu}(\vec{x}) =
 -\frac{2T_0}{\sqrt{-g_4}} \frac{\delta \ln Z}{\delta g^{\mu\nu}(\vec{x})}\\
& J^\mu (\vec{x}) = \frac{T_0}{\sqrt{-g_4}} \frac{\delta \ln Z}{\delta \mathcal{A}_\mu (\vec{x})}\ .
\end{split}
\end{align}
When regarding the partition function as a functional of:
\begin{equation}\label{anom:indep_var}
\ln Z \equiv W(e^\sigma, A_0,a_i, A_i,g^{ij}, T_0,\mu_0),
\end{equation}
these can be recast as:
\begin{gather}
\begin{split}
T_{00} (\vec{x}) =  & -\frac{T_0 e^{2\sigma (\vec{x}) }}{\sqrt{-g_4}} \frac{\delta W}{\delta \sigma (\vec{x}) }\ ,
\end{split} \qquad
\begin{split}
T_0^i (\vec{x}) = &\frac{T_0}{\sqrt{-g_4}} \left(\frac{\delta W}{\delta a_i(\vec{x})} - A_0 (\vec{x}) \frac{\delta W}{\delta A_i (\vec{x})} \right)\ ,\notag
\end{split} \\
\begin{split}\label{anom:JT_formulas}
T^{ij} (\vec{x})  = & -\frac{2T_0}{\sqrt{-g_4}} g^{il} g^{jm} \frac{\delta W}{\delta g^{lm} (\vec{x})}\ ,
\end{split}\\ \notag
\begin{split}
J_0  (\vec{x}) =& -\frac{T_0 e^{2\sigma (\vec{x})}}{\sqrt{-g_4}} \frac{\delta W}{\delta A_0 (\vec{x})}\ ,
\end{split}\qquad
\begin{split}
J^i (\vec{x}) = &\frac{T_0}{\sqrt{-g_4}} \frac{\delta W}{\delta A_i (\vec{x})}\ .
\end{split}
\end{gather}
Note that the formulas are preferably presented with \emph{upper spatial} and \emph{lower temporal} indexes. This is due to the fact that tensors with such an index structure are invariant under Kaluza-Klein gauge transformations ($t\rightarrow t +\phi(\vec{x})$, $a_i\rightarrow a_i -\del_i \phi(\vec{x})$).

Plugging the most general partition function for a 3+1 dimensional charged fluid (equation \eqref{anom:partition}) into the relations \eqref{anom:JT_formulas}, the authors of \cite{Banerjee:2012iz} found the following results for the stress-energy tensor and charge current:
\begin{align}
\begin{split}\label{anom:T00}
T_{00} = - e^{2\sigma}\left(P-aP_a-bP_b\right), \qquad T^{ij}=Pg^{ij},
\end{split}\\
\begin{split}\label{anom:T0i}
T_0^i = e^{-\sigma} \epsilon^{ijk} \left[\left(-\frac{1}{2}CA_0^2+C_2 T_0^2\right)\del_j A_k
            -\left(\frac{C}{6}A_0^3+C_2A_0T_0^2\right) \del_j a_k \right]
\end{split}\\
\begin{split}
J_0 = -e^{\sigma} P_b-e^\sigma \epsilon^{ijk} \left[\frac{C}{3}A_i \del_j A_k+\frac{C}{3}A_0A_i\del_j a_k \right],
\end{split}\\
\begin{split}\label{anom:Ji}
J^i = e^{-\sigma}\epsilon^{ijk} \left[\frac{2}{3}CA_0\del_j A_k + \left(\frac{C}{6}A_0^2+C_2T_0^2\right)\del_j a_k + \frac{C}{3}A_k \del_j A_0\right]
\end{split}
\end{align}
where $a\equiv e^{-\sigma}T_0$, $b\equiv e^{-\sigma}A_0$, and $P_a, P_b$ are the partial derivatives of $P$ with respect to $a$ and $b$ respectively. Some $T_0$ factors were missing in equation (3.9) of \cite{Banerjee:2012iz} and are added here.

The covariant form of the current:
\begin{equation}\label{anom:cov_def}
    \tilde J^\mu = J^\mu - \frac{C}{6}\epsilon^{\mu\nu\rho\sigma}\mathcal{A}{}_\nu \mathcal{F}{}_{\rho\sigma}\ ,
\end{equation}
is given by:
\begin{align}\label{anom:JiCov}
\begin{split}
\tilde J_0 = -e^{\sigma} P_b; \qquad
\tilde J^i = e^{-\sigma}\epsilon^{ijk} \left[CA_0\del_j A_k + \left(\frac{C}{2}A_0^2+C_2T_0^2\right)\del_j a_k \right]\ .
\end{split}
\end{align}

Using the metric and gauge field dependence of the stress-energy tensor and the charge current, which is fully revealed in Eqs.~\eqref{anom:T00}-\eqref{anom:JiCov}, it is straightforward to find Kubo formulas for the thermal constants $C$ and $C_2$.
In this case $C$ and $C_2$ must be constants rather than functions of the temperature and chemical potential in order for the partition function to have the required anomaly and invariance properties. One needs now to vary the stress-energy tensor and charge current with respect to the appropriate component of the metric/gauge-field to get the retarded correlation functions that constitute the Kubo formulas for the thermal constants.

Since we work with a stationary background, the Kubo relations we shall find will only allow us to determine the thermal non-dissipative
transport coefficients, i.e. those coefficients that affect the fluid dynamics at equilibrium. These will be determined by the correlation functions evaluated at zero frequency. Up to powers of $i$, zero frequency retarded correlators equal zero frequency Euclidean correlators. Equivalently, the Kubo relations can be worked out directly in Euclidean space as in \cite{Moore:2012tc}, relating the thermal constants and, as a consequence, the non-dissipative transport coefficient, to Euclidean correlation functions.

It should be noted that if we wish to keep the independent variables as in \eqref{anom:indep_var}, i.e. $e^\sigma, A_0,a_i, A_i,g^{ij}$, when varying w.r.t the gauge field and metric perturbation, we must vary according to equation \eqref{anom:JT_formulas} type formulas. Special attention must be paid when raising/lowering stress-tensor/charge-current indexes, since these operations normally involve extra metric factors and as a consequence do not in general commute with a variation w.r.t the metric. Equivalently, one can translate back $e^\sigma, A_0,a_i, A_i,g^{ij}$ into the original gauge field and metric variations $\delta \mathcal{A}_\mu, \delta g_{\mu\nu}$. The variation needed to obtain Kubo-formula is then immediate. We will be using both methods alternately depending on which is simpler for the case studied. For the second order fluid studied in appendix \ref{sec:second_order} for example, corrections from raising/lowering indexes using the set of variables $e^\sigma, A_0,a_i, A_i,g^{ij}$ become involved, so the second method is preferable. For the cases studied in this section and the next however, this set of variables will suffice.

We will be using the following definition for the Green function:
\begin{align}
\begin{split}\label{anom:Green_def}
&G^{(\mu_1\nu_1),\dots, (\mu_n\nu_n), \rho_1,\dots,\rho_k}
 \left(p_1,\dots,p_{n+k-1},-\Sigma_1^{n+k-1} p_i\right)
\\ &
\qquad\qquad
 =  \int d^4 x_1 \dots d^4 x_{n+k-1} \cdot e^{-i \sum_1^{n+k-1} p_i \cdot x_i} \times
\\&
\qquad\qquad\qquad\qquad\quad
\times \frac{ 2^n \cdot T_0 \cdot \del^{n+k} \ln Z}
{\del g_{\mu_1\nu_1}(x_1)  \dots\del g_{\mu_n\nu_n} (x_n)\del \mathcal{A}_{\rho_1} (x_{n+1}) \dots \del \mathcal{A}_{\rho_k}(0)} \biggr|_{
\delta \mathcal{A_\mu}=0  \atop \delta g_{\mu\nu}=0 }
\\ &
\qquad\qquad
=
 \langle T^{\mu_1\nu_1}(p_{1})  \dots T^{\mu_n\nu_n}(p_{n}) J^{\rho_1}(p_{n+1}) \cdots J^{\rho_k}(-\Sigma_1^{n+k-1} p_i) \rangle +\mbox{c.t.}
\end{split}
\end{align}
This is very similar to the Euclidean n-point function defined in \cite{Moore:2012tc}, with a small difference, we differentiate w.r.t the ``Lorentzian'' metric, which is a factor of $i$ different for each $t$ index compared to the definition in \cite{Moore:2012tc}. Otherwise, the definitions are the same (for a Lorentzian definition see \cite{Moore:2010bu}).
To evaluate this type of Green functions using Feynman diagrams (cf. \cite{Landsteiner:2011cp}), one passes to Euclidean space. We therefore find it advantageous to work with this Euclidean definition all along (up to the above mentioned factors of $i$).

Our definition for the Green function \eqref{anom:Green_def} involves multiple metric/gauge-field derivatives acting on the  partition function of our system. This partition function can be thought of as the Euclidean action of the system with the metric given in \eqref{anom:eqMetric} and with time coordinate compactified to a circle of length $1/T_0$. Since the system is stationary, we are allowed to replace time integration with $1/T_0$ factor and time functional derivative with a $T_0$ factor. We can thus content ourselves with 3-integration and 3-differentiation in equation \eqref{anom:Green_def}. One extra $T_0$ factor is present since we have one extra differentiation.
The first differentiation stage was already performed in Eqs~\eqref{anom:T00}-\eqref{anom:JiCov}), which we will use.

Two comments are in order. First, when repeatedly differentiating the energy functional, each derivative can either pull an extra factor of $T^{\mu\nu}/J^\mu$ or it can act on a factor of $T^{\mu\nu}/J^\mu$, already pulled down by the previous $g_{\mu\nu}/\mathcal{A}_\mu$ derivatives. This is the origin of the contact terms (c.t.) on the last line of equation \eqref{anom:Green_def}.
Second, what we get by differentiation in the intermediate steps is not really the stress tensor $T^{\mu\nu}$ but rather the stress-energy tensor density $\sqrt{-g_4}T^{\mu\nu}$. One can check that when evaluated in flat space, none of the Kubo formulas presented in this paper change due to the additional contact terms implied by the differentiation of the extra $\sqrt{-g_4}$ factor.

\subsection{Extracting the Kubo Relations}
Let us start by varying $T^{0j}$ with respect to the $i$-th component of the gauge field $\mathcal{A}_i$. Using the set of variables from equation \eqref{anom:indep_var}, this would mean varying $T^{0j} = (T_0^j-g_{0k}T^{kj})/g_{00}$ with respect to the Kaluza-Klein gauge invariant ``gauge field'' $A_i$. Since $T^{kj}$ does not depend on $A_i$, (and neither does $g_{00}$), upon setting the metric and gauge field perturbation to zero we obtain (in momentum space):
\begin{align}
\begin{split}\label{anom:Kubo rel_TJ}
G^{i,0j}(k,-k)\biggr|_{\omega=0} = - \frac{i}{2} \epsilon^{ijk} k_k \left(C\mu^2-2C_2 T^2\right) + O(k^2).\\
\end{split}
\end{align}
$G$ is the Euclidean Green function of stress tensors and currents \eqref{anom:Green_def} evaluated in flat space.
The zero frequency limit removes any dissipative contribution which might not be accounted for by our equilibrium partition function analysis.
Since we have set the metric and gauge field perturbation to zero, $T=T_0$ is the equilibrium temperature.
Similarly $\mu=\mu_0=A_0$ is the equilibrium chemical potential.

We have thus obtained a Kubo relation for $C_2$:
\begin{equation}\label{anom:C_2}
\boxed{C_2=\frac{C}{2}\left(\frac{\mu}{T}\right)^2-\frac{i}{2T^2}\lim_{k_n\rightarrow 0} \sum_{ij}  \frac{\epsilon_{ijn}}{k_n} G^{i,0j}(k_n,-k_n)\biggr|_{\omega=0}}\ ,
\end{equation}
where $k_n$ is the external momentum and $C$ is the chiral anomaly coefficient. The identification of $C$ with the anomaly coefficient can be inferred from the expected transformation properties of the equilibrium partition function under gauge transformation. Alternatively, one can vary the divergence of the current \eqref{anom:Ji} twice, with respect to both $A_0$ and $A_k$, restoring the anomaly non conservation equation. Note, that \eqref{anom:Ji} is the consistent form of the current.

Similar Kubo relations follow from varying the current $J^j$ given in equation \eqref{anom:Ji} (or its covariant counterpart $\tilde{J}^j$ given in equation \eqref{anom:JiCov}) with respect to $i$-th component of the gauge field $\mathcal{A}_i$:
\begin{align}
\begin{split}\label{anom:Kubo rel_JJ}
&G^{i,j}(k,-k)\biggr|_{\omega=0} = -\frac{2}{3} \epsilon^{ijk} i k_k C \mu + O(k^2)\ ,
\quad\qquad
\widetilde G^{i,j}(k,-k)\biggr|_{\omega=0} =- \epsilon^{ijk} i k_k C \mu + O(k^2)\ ,\\
\end{split}
\end{align}
where we have again used the set of independent variables \eqref{anom:indep_var} when varying. $\widetilde G$ refers to a correlator that is obtained from the variation of the covariant current, $\widetilde G^{i,j} = \delta \tilde J^j / \delta \mathcal A_i$ (rather than the consistent current as in \eqref{anom:Green_def}). This is usually the type of Green functions obtained in hydrodynamic analysis of Kubo-relations (see e.g.\cite{Amado:2011zx}).

Some more Kubo formulas can be obtained by varying the stress tensor $T^{0j}$ w.r.t the metric component $g_{0i}$ (for $i\neq j$). Upon setting the metric and gauge field perturbation to zero we get:
\begin{align}
\begin{split}\label{anom:Kubo rel_TT}
& G^{0i,0j}(k,-k)\biggr|_{\omega=0} = - \epsilon^{ijk}ik_k(\frac{1}{3}C\mu^3-2C_2\mu T^2) + O(k^2),
\end{split}
\end{align}
where we have used $T^{0j} = (T_0^j-g_{0k}T^{kj})/g_{00}$, and the variation w.r.t $g_{0i}$ was performed using the set of independent variables of equation \eqref{anom:indep_var}, according to equation \eqref{anom:JT_formulas} type differentiation rules:
\begin{align}
\frac{\delta}{\delta g_{0i}}=\frac{1}{g_{00}} \left( \left[\frac{\delta}{\delta a_i} - A_0 \frac{\delta}{\delta A_i} \right] -g_{k0}\left[-2g^{kl}g^{im}\frac{\delta}{\delta g^{lm}}\right]\right).
\end{align}

\subsection{Hydrodynamic Transport Coefficients}
The most general allowed form for the hydrodynamic stress-energy tensor and charge current can be found on symmetry grounds to be:
\begin{align}
\begin{split}\label{anom:Tmunu}
T^{\mu\nu}=(\epsilon+P)u^\mu u^\nu +P g^{\mu\nu} - \eta \sigma^{\mu\nu} -\zeta \nabla \cdot u P^{\mu\nu},
\end{split}\\
\begin{split}\label{anom:Jmu}
J^\mu=qu^\mu + \sigma\left(E^\mu-T P ^{\mu\alpha}\del_\alpha \left(\frac{\mu}{T}\right)\right)+
\xi_\omega\omega^\mu +\xi_B B^\mu,
\end{split}
\end{align}
with $\epsilon$ the energy density, $P$ the pressure, $q$ the charge density, $s$ (which we will use later) the entropy density, $u^\mu$ the normalized ($u^\mu u_\mu=-1$) fluid  velocity, $P^{\mu\nu} = g^{\mu\nu}+u^\mu u^\nu$ the transverse projector, $\sigma^{\mu\nu} = P^{\mu\alpha}P^{\nu\beta} \left(\frac{\nabla_\alpha u_\beta+\nabla_\beta u_\alpha}{2}-\frac{\nabla_\alpha u^\alpha}{3} P^{\mu\nu}\right)$ the shear tensor, $\omega^\mu = \frac{1}{2} \epsilon^{\mu\nu\rho\sigma} u_\nu \del_\rho u_\sigma$ the vorticity vector, $E^\mu = \mathcal{F}^{\mu\nu} u_\nu$ the electric field and $B^\mu = \frac{1}{2} \epsilon^{\mu\nu\rho\sigma} u_\nu \mathcal{F}_{\rho\sigma}$ the magnetic field. All the hydrodynamic expressions in this paper will be presented in the Landau frame, i.e. the frame in which the stress-energy tensor and current corrections are transverse to the fluid velocity.

One can then write the most general equilibrium solution for the fluid fields ($T,\mu,u^\mu$) as a function of the external fields. The zeroth order solution consists of the local red shifted values:
\begin{equation}
\begin{split} \label{anom:zeroth_sol}
\hat T =& \ T_0 e^ {-\sigma}\ ,\\
\hat \mu = & \ A_0 e^{-\sigma}\ ,\\
\hat u^\mu = & \ (1,0,0,0)e^{-\sigma} \ .
\end{split}
\end{equation}
The first order solution consists of any addition to the above ($\delta T,\delta\mu,\delta u^\mu$) which is allowed by symmetry and is of first order in derivatives of the external sources. Plugging these into the stress-energy tensor and charge current  \eqref{anom:Tmunu}-\eqref{anom:Jmu} and evaluating them on the equilibrium configuration \eqref{anom:eqMetric}-\eqref{anom:eqGauge} one obtains a general expression for the stress-energy  tensor as a function of the external background fields.

Comparing this form with the stress-energy tensor and current obtained by varying the equilibrium partition function \eqref{anom:T00}-\eqref{anom:JiCov}, one can express the non-dissipative hydrodynamic transport coefficients $\xi_\omega, \xi_B$ in terms of the partition function constants $C_2,C$ \cite{Banerjee:2012iz}:
\begin{align}
\begin{split}\label{anom:vort_cond}
\xi_\omega = C\mu^2 - 2C_2 T^2 - \frac{q}{\epsilon+P}\left(\frac{2}{3}C\mu^3- 4C_2T^2\mu\right),
\end{split}\\
\begin{split}\label{anom:magn_cond}
\xi_B = C\mu - \frac{q}{\epsilon+P}\left(\frac{1}{2}C\mu^2- C_2T^2\right).
\end{split}
\end{align}

Expressing these as Kubo formulas for the chiral transport coefficients using \eqref{anom:Kubo rel_TJ}-\eqref{anom:Kubo rel_TT}:
\begin{align}
\begin{split}\label{anom:vort_kubo}
\boxed{ \xi_\omega = \lim_{k_n\rightarrow 0} \sum_{ij} \frac{i}{k_n} \epsilon_{ijn} \left[ G^{i,0j}(k_n,-k_n)  - \frac{q}{\epsilon+P}
  G^{0i,0j}(k_n,-k_n) \right] \biggr|_{\omega=0} }\ ,
\end{split}\\
\begin{split}\label{anom:magn_kubo}
\boxed{ \xi_B =
 \lim_{k_n\rightarrow 0} \sum_{ij} \frac{i}{2k_n}\epsilon_{ijn}
    \left[ \widetilde G^{ i,}{}^j(k_n,-k_n)  - \frac{q}{\epsilon+P}   G^{i, 0j}(k_n,-k_n)  \right]\biggr|_{\omega=0} }\ ,
\end{split}
\end{align}
we reproduce the Kubo formulas of \cite{Amado:2011zx}.\footnote{Up to a minus sign which appears due to a difference $ k\leftrightarrow- k$ in their definition of Green functions, and a factor 1/2 difference in the vortical conductivity due to a factor of 1/2 difference in the definition of the vorticity.}
When equating the hydrodynamic stress-energy tensor on the most general equilibrium fluid solution to the one derived from the equilibrium partition function one in fact solves for the fluid profile in this very special equilibrium case.
The $k_n\rightarrow 0$ limit is taken in order to get rid of terms of higher order in derivatives.
For an example of how to evaluate these formulas see \cite{Landsteiner:2012kd}. 

\section{Superfluid Dynamics in 3+1 Dimensions}\label{sec:supfl}
In this section we derive new Kubo formulas for the non-dissipative transport coefficients associated with the flow of a 3+1 dimensional relativistic superfluid up to first order in the derivative expansion. The authors of \cite{Bhattacharya:2011tra} found that for time reversal invariant superfluids all the transport coefficients can be expressed in terms of fourteen independent functions in the parity even sector, and six independent functions in the parity odd sector.
All the parity even functions and one of the parity odd functions are dissipative in the sense that they result in entropy production.
We are therefore left with five parity odd entropically non-dissipative independent functions.
Of these, only two ($\sigma_8$ and $\sigma_{10}$ in the notations of  \cite{Bhattacharya:2011tra})
multiply terms that do not vanish at equilibrium. These two functions (and their derivatives) can be used to express \emph{all} the (thirteen) superfluid transport coefficients that affect the superfluid dynamics in equilibrium.
In the absence of time reversal invariance three more thermal functions are needed to express all the non-dissipative superfluid transport coefficients.

We require our superfluid to be neither parity preserving nor time reversal invariant. However, we require that our fluid is CPT invariant. Our analysis is divided into two parts. In the first part we derive Kubo formulas for the parity even transport coefficients. In the second we analyze the parity odd transport coefficients. The Kubo formulas for each sector (even/odd) receive no mixed contribution from the other sector, as will be shown throughout the analysis. Therefore the study could have been carried out separately for each sector. In the discussion we draw general conclusions from the Kubo analysis about the nature of the parity odd superfluid transport coefficients. We also present new identities that are revealed when performing the analysis.

\subsection{Preliminaries I - Superfluid Hydrodynamics}\label{subsec:supfl:prelimI}
A superfluid is the fluid phase of a system with a spontaneously broken global symmetry. For `s' wave superfluids the symmetry breaking manifests itself in the appearance of a vacuum expectation value of a charged scalar operator. The phase of the charge condensate induces a new massless Goldstone mode into the theory. Being massless the Goldstone mode participates in the hydrodynamics. The motion of a superfluid consists of two distinct flows. The first is the flow of the normal part of the fluid which is encoded in the fluid velocity $u^\mu$. The second is the flow associated with the condensate (superfluid) part. This part has a velocity in the direction of the gradient of the Goldstone phase. When considering a background gauge field $\mathcal{A}_\mu$ as well, it is the covariant derivative of the Goldstone phase $\phi$ that points in the direction of the superfluid velocity and thus enters the hydrodynamic description of the system:
\begin{equation} \label{supfl:xi}
\xi_\mu \equiv -\del_\mu \phi+\mathcal{A}_\mu\ .
\end{equation}
The superfluid velocity is then given by $u^\mu_s = -\xi^\mu / \xi$, where $\xi = \sqrt{-\xi^\mu \xi_\mu}$.
The eight variables of superfluid hydrodynamics are $u^\mu(x)$, $\xi^\mu(x)$ and $T(x)$.

It is sometimes convenient to replace these eight fields by nine hydrodynamic fields subject to a single constraint. The additional field in that description is the local chemical potential $\mu(x)$ related to the other fields by the ``Josephson equation'':
\begin{equation}
u(x) \cdot \xi(x)= \mu(x)+ \mu_{diss}(x)
\end{equation}
where $\mu_{diss}(x)$ is a function of derivatives of the fluid variables. At zeroth order in the derivative expansion this relation simply equates the component of the `generalized' gauge field $\xi^\mu$ in the direction of the normal-fluid velocity with the chemical potential $\mu$. It is sometimes convenient to use the definition:
\begin{equation}\label{supfl:zeta_def}
\zeta^{\mu} = P^{\mu\nu} \xi_\nu\ ,
\end{equation}
for the component of the $\xi^\mu$ orthogonal to $u^\mu$.

The equations of superfluid dynamics are:
\begin{equation}
\begin{split}\label{supfl:cons_eqs}
\del_\mu T^{\mu\nu} = \mathcal{F}^{\nu\mu}J_{\mu}\ ,\\
\del_\mu J^\mu = cE_\mu B^\mu\ ,\\
\del_\mu \xi_\nu-\del_\nu \xi_\mu = \mathcal{F}_{\mu\nu}\ ,
\end{split}
\end{equation}
where the stress tensor and current are given by:
\begin{equation}
\begin{split}\label{supfl:constit0}
T^{\mu\nu} & = (\epsilon+P)u^\mu u^\nu + P \eta^{\mu\nu} + f \xi^\mu \xi^\nu + \pi^{\mu\nu}\ ,\\
J^{\mu} & = q u^\mu -f \xi^\mu + j^{\mu}_{diss}\ .
\end{split}
\end{equation}
The superfluid constitutive relations are expressions for $\pi^{\mu\nu}$, $j^{\mu}_{diss}$ and $\mu_{diss}$ in terms of derivatives of the superfluid dynamical fields ($u^\mu,\xi^\mu,T,\mu$) and background fields (metric/gauge field).
All the thermal coefficients in equation \eqref{supfl:constit0} are functions of the three scalars: $T,\mu,\xi$.
They are not independent but rather given in terms of a single thermodynamical pressure function $P(T,\mu,\xi)$ through the thermodynamic relations:
\begin{equation}
\begin{split}\label{supfl:Gibbs_Duhem}
\epsilon+P &\ = sT+q\mu\ , \\
dP = &\ sdT+qd\mu+\frac{1}{2}fd\xi^2\ .
\end{split}
\end{equation}

The equations of superfluid dynamics change their detailed form under field redefinitions. The temperature, chemical potential and (normal) fluid velocity field, are only well defined at the zeroth order in the derivative expansion. At higher orders in derivatives they are ambiguous. This means that a redefinition $u^\mu\rightarrow u^\mu+\delta u^\mu, T \rightarrow T+\delta T, \mu \rightarrow\mu+\delta\mu$ accompanied by an appropriate adaptation of the constitutive relations can provide an equivalent description of superfluid dynamics. This is not true for the Goldstone phase gradient $\xi^\mu(x)$ which is microscopically well defined. To completely fix the equations of superfluid dynamics we therefore need to specify a `frame' (that is, a non ambiguous definition of $u^\mu, T$ and $\mu$). This is achieved by specifying certain conditions on the derivative corrections to the constitutive relations (i.e. on $\pi^{\mu\nu}, j_{diss}^{\mu}$ and $\mu_{diss}$). For example the `Transverse Frame' is defined to be the frame in which:
\begin{equation}
\begin{split}\label{supfl:frame_fix}
\pi^{\mu\nu} u_{\nu} = 0\ ,\\
j_{diss}^{\mu} u_{\mu} = 0\ .
\end{split}
\end{equation}

As mentioned above, the superfluid constitutive relations are expressions for the derivative corrections to the stress tensor $\pi^{\mu\nu}$, charge current $j_{diss}^{\mu}$ and chemical potential $\mu_{diss}$ in terms of derivatives of the fluid dynamical fields (and background data).
It is sometimes convenient to specify the constitutive relations in terms of field redefinition invariant combinations of $\pi^{\mu\nu}$ and $j_{diss}^{\mu}$ and $\mu_{diss}$ instead of specifying the full $\pi^{\mu\nu}$ and $j_{diss}^{\mu}$ and $\mu_{diss}$ in a specific frame. In such a case the full constitutive relations are completely determined after adding five frame fixing conditions such as \eqref{supfl:frame_fix}.

It is possible to obtain various constraints on the most general form allowed for the constitutive relations by requiring the existence of an entropy current of positive divergence. This was done in \cite{Bhattacharya:2011eea} for parity preserving time reversal invariant superfluids, in \cite{Bhattacharya:2011tra} for parity non-preserving (but still time reversal invariant) superfluids, in \cite{Neiman:2011mj} for the parity odd sector of superfluids with multiple unbroken charges, and finally in subsection (3.1) of \cite{Bhattacharyya:2012xi} for a single charge without assuming parity/time reversal invariance. We will present some of these results in the following sections where needed.

\subsection{Preliminaries II - Superfluid Effective Action}\label{subsec:supfl:prelimII}

The partition function analysis for superfluids was carried out in \cite{Bhattacharyya:2012xi} with one major difference compared to the partition function analysis of \cite{Banerjee:2012iz} that we used in the previous section.
Instead of considering the partition function for superfluids as a function of the external sources ($\mathcal{A}_\mu, g_{\mu\nu}$),  the local effective action for the Goldstone phase gradient was used.
To get from the local effective action to the full partition function one has to integrate out the Goldstone boson. In the classical limit this amounts to solving the equation of motion for the Goldstone mode and plugging back the solution into the effective action. We will be working in this limit.

The theory admits a degenerate set of vacua which break spontaneously the symmetry,
the Goldstone mode is an excitation along these vacua. 
Integrating out this massless mode therefore results in a highly non-local expression for the partition function as a function of the external fields. It is therefore easier to use the effective action for the Goldstone phase gradient directly to derive the stress tensor and current instead of using the full partition function, integrating out the Goldstone phase at the last step of the calculation. This has the advantage that the Goldstone phase can be treated as independent of the external sources at the step in which the stress tensor and charge current are obtained by differentiation. One therefore doesn't have to deal with solving the equation of motion for the Goldstone phase and recovering its exact dependence on the background fields. This is no longer true
when computing higher correlation functions, since they are not just determined by the variation of the action evaluated at the solution, but can also receive contributions from the variation of the solution itself.

In our Kubo formula derivation we will use the results of \cite{Bhattacharyya:2012xi} for the stress tensor and current obtained as explained above.
We will also solve the Goldstone equation of motion (minimize the effective action) to find the expectation value of the Goldstone field in the classical limit.
We will then vary these quantities with respect to the external background sources to obtain Kubo formulas for the transport coefficients.
We will have to pay careful attention to the variation of the Goldstone solution $\zeta_{eq}$ w.r.t the external sources, because of the corrections induced by the variation of the Goldstone solution.

When comparing the hydrodynamic stress-energy tensor and charge current with the ones obtained from the effective action, the authors of \cite{Bhattacharyya:2012xi} regarded the equilibrium solution for the Goldstone phase gradient as independent of the other background fields. This is due to the non-locality of the classical solution, which lead to the conclusion that cancelations between the Goldstone phase gradient and other local functionals of the background fields are impossible, except for those implied by the equation of motion of the Goldstone phase gradient.

Since Kubo formulas are eventually evaluated on a flat background with constant gauge fields, at the final stage of our analysis, after setting the sources to zero, the solution for the gradient of the Goldstone phase becomes a constant independent thermal equilibrium parameter.
We will denote the component of the equilibrium Goldstone phase gradient in the direction perpendicular to the normal fluid velocity in the absence of sources $\zeta^i_{0}$.
The addition of the equilibrium Goldstone phase gradient strongly resembles the addition of a finite chemical potential to the normal fluid. In the absence of sources we therefore set $\mathcal A^\mu = (\mu_0,\zeta_0^i)$.

As we mentioned in the first part of those preliminaries, an equilibrium solution for superfluid dynamics in flat space is fixed by eight thermal parameters. The general form we were using for the metric \eqref{anom:eqMetric} made use of the
coordinate freedom to fix the alignment of the time-like killing vector with the $t$ coordinate. This alignment fixes the equilibrium velocity in the absence of sources to be $u^\mu=(1,0,0,0)$. This still leaves us with five free parameters ($T_0$, $\mu_0$  and $\zeta^i_0$) at thermal equilibrium in flat space.
In a stationary setup these are the values that the temperature, chemical potential (zero component of the gauge field) and spatial components of the Goldstone phase gradient will obtain after setting the sources to zero. They are all constants.
When evaluating the Kubo formulas, that will be presented in the next subsection, in terms of Euclidean (flat space) thermal QFT Feynman diagrams we expect a change in the fermion propagators of the form $i \omega \rightarrow i \omega_n +\mu_0$, $q^i \rightarrow q^i +\zeta^i_0$, where $\omega_n$ are the Matsubara frequencies and $q$ is the spatial momenta of the fermion line. This should be accompanied by an appropriate change of the stress-tensor/charge-current vertex operators (to account for the superfluid contribution). 

Every part of our analysis will be carried out in two steps.
First, only parity even contributions to the effective action and superfluid constitutive relations will be considered. Kubo formulas for the parity even thermal functions $c_1,c_2,c_3$ of \cite{Bhattacharyya:2012xi} will be presented along with the associated transport coefficients.
Time reversal invariance is not assumed.
In the second step we will consider the parity odd sector. Kubo formulas for the parity odd thermal functions $g_1,g_2$ and the associated transport coefficients will be presented.
In the discussion we present conclusions drawn from the Kubo analysis about the nature of the parity odd superfluid transport coefficients.

We start by presenting parts of the effective action analysis of \cite{Bhattacharyya:2012xi} that we will need to derive the Kubo formulas for both the parity even and parity odd transport coefficients.

\subsubsection{Parity Even Effective Action}
The most general parity even equilibrium effective action one can build from the Goldstone phase gradient and external sources up to first order in derivatives (keeping Kaluza Klein, gauge and 3d diff-invariance intact) is given by:
\begin{equation}
\begin{split}\label{supfl:part_even}
S &\ = S_0 + S_1^{even}\ ,\\
S_0&\ = \int d^3 x \sqrt{g_3} \frac{1}{\hat T} P(\hat T, \hat \mu , \xi^2)\ ,\\
S_1^{even}&\  = \int d^3 x \sqrt{g_3}\  f \left[c_1 (\zeta \cdot \del) \hat T
+ c_2 (\zeta \cdot \del) \hat \nu
+ c_3 (\zeta \cdot \del) \zeta^2 \right]\ ,
\end{split}
\end{equation}
where the background metric and gauge field were defined in \eqref{anom:eqMetric}-\eqref{anom:eqGauge} and \eqref{anom:KK_inv_gauge}, $\hat T$ and $\hat \mu$ where defined in \eqref{anom:zeroth_sol},
\begin{equation}
\hat \nu \equiv \frac{\hat \mu}{\hat T} = \frac{A_0}{T_0}\ ,
\end{equation}
$\xi_\mu$ is the superfluid phase gradient of \eqref{supfl:xi} and,
\begin{equation}
\zeta_i \equiv \xi_i -a_i A_0 = -\del_i \phi + A_i
\end{equation}
is the Kaluza Klein gauge invariant combination of the superfluid phase gradient.
By convention $\zeta_i$'s index is raised and lowered with the three dimensional metric $g_{ij}$. On the zeroth order solution \eqref{anom:zeroth_sol} the above $\zeta_i$ indeed turns out to be the orthogonal component of $\xi^\mu$ as implied by \eqref{supfl:zeta_def}. We therefore use the same symbol for these two quantities. All the functions $c_i$ are given in terms of the independent variables:
\begin{equation}
c_i = c_i (\hat T, \hat \nu, \zeta^2)\ .
\end{equation}
$f$ is defined through \eqref{supfl:Gibbs_Duhem}:
\begin{equation}
f = 2 \frac{\del P}{\del \xi^2} = - 2 \frac{\del P}{\del \zeta^2}\ ,
\end{equation}
where the differentiation with respect to $\zeta^2$ is carried out at constant $\hat T$ and $\hat \nu$, after the appropriate change of variables. After comparing the hydrodynamic stress-energy tensor and current to the ones derived from the effective action at zeroth order, it can be demonstrated that $P$ from \eqref{supfl:part_even} and \eqref{supfl:Gibbs_Duhem} are the same thermal pressure function.

The leading derivative order equation of motion for the Goldstone phase can be obtained by varying the action $S_0$ with respect to $\phi$ and is given by\cite{Bhattacharyya:2012xi}:
\begin{equation}\label{supfl:zero_eom}
\nabla_i \left(\frac{f}{T}\zeta^i\right) = 0,
\end{equation}
where the derivative is covariant with respect to the 3 dimensional spatial metric $g_{ij}$ (the next order corrections to the equation of motion originating from $S_1$ are given in appendix \ref{app:Goldstone_eom_corrections}).
 We will denote the solution to this equation $\phi_{eq}$, and the associated $\vec\zeta$ will be denoted $\vec\zeta_{eq}$. It will be in general a functional of the external sources. In the classical limit $\phi_{eq}$ is the expectation value of the Goldstone phase.

The solution to the equation of motion (including the appendix \ref{app:Goldstone_eom_corrections} corrections) at linear order in the sources is given in appendix \ref{app:solving_eom}.
In momentum space, we have for the special case of $\vec\zeta_0 \perp \vec k$ and $\delta g^{ij}=0$:
\begin{equation}
\begin{split}\label{supfl:solution_eom}
 \langle \phi \rangle_{eq} = & -i \frac{k\cdot \delta A}{k^2}
 - 2 (\zeta_0 \cdot \delta A) T_0c_3
     + \sigma c_1T_0^2
     -\delta A_0  c_2 + O(\delta^2, k)\ ,
\end{split}
\end{equation}
where we used the following definitions $\delta A_0 \equiv A_0-\mu_0$, $\delta A_i \equiv A_i-(\zeta_0)_i$ and
$\delta^2$ stands for any contribution which is of second order in the variation of the sources.
For $\vec\zeta_0 \parallel \vec k$ we have:
\begin{equation}
\begin{split}\label{supfl:solution_eom}
 \langle \phi \rangle_{eq} = & -i \frac{k\cdot \delta A}{k^2}\ .
\end{split}
\end{equation}
Using these to express the transverse superfluid velocity in momentum space gives:
\begin{equation}
\begin{split}\label{supfl:perp_sol_eom}
 \zeta_{eq}^i = A^i - i k^i  \langle \phi \rangle_{eq}\ ,
\end{split}
\end{equation}
where in the absence of background source variation $A^i=\zeta_0^i$.

The stress-tensor and charge-current are obtained by varying the effective action with respect to the various sources according to Eqs.~\eqref{anom:JT_formulas}. For our analysis we will only need $J^i$ and $T_0^i$. We list their explicit expressions as given in \cite{Bhattacharyya:2012xi}:\footnote{The conversion between the $c_i$'s and the $f_i$'s of \cite{Bhattacharyya:2012xi} is given by:
\begin{equation}
c_1 \equiv \frac{f_1}{fT}+\frac{1}{T}\frac{\del f_3}{\del \hat T};
\qquad
c_2 \equiv \frac{f_2}{fT}+\frac{1}{T}\frac{\del f_3}{\del \hat \nu};
\qquad
c_3 \equiv \frac{1}{T}\frac{\del f_3}{\del \zeta^2};
\end{equation}}
\begin{align}
\begin{split}\label{supfl:even_1_current}
J^{i} = & \frac{\hat T}{\sqrt{g_3}} \frac{\del S}{\del A_i} = -f\zeta_{eq}^i
+ f\hat T g^{ij} (c_1 \del_j \hat T + c_2 \del_j \hat \nu +c_3 \del_j \zeta_{eq}^2)
- 2 \zeta_{eq}^i  f \zeta_{eq} \cdot \del (c_3 \hat T)
\\
& \ + 2 \zeta_{eq}^i \hat T
\left[ \frac{\del (f c_1)}{\del \zeta_{eq}^2} \zeta_{eq} \cdot \del \hat T + \frac{\del (f c_2)}{\del \zeta_{eq}^2} \zeta_{eq} \cdot \del \hat \nu +\frac{\del (f c_3)}{\del \zeta_{eq}^2} \zeta_{eq} \cdot \del \zeta_{eq}^2
 \right]
\end{split}\\
\begin{split}\label{supfl:even_1_stress}
T_0^{i} & =  \frac{\hat T}{\sqrt{g_3}} \left[ \frac{\del S}{\del a_i} - A_0 \frac{\del S}{\del A_i} \right]
= - A_0 J^i =
\\ &
  fA_0\zeta_{eq}^i
- A_0 f\hat T g^{ij} (c_1 \del_j \hat T + c_2 \del_j \hat \nu +c_3 \del_j \zeta_{eq}^2)+ 2 \zeta_{eq}^i  A_0f \zeta_{eq} \cdot \del (c_3 \hat T)
\\&
- 2 \zeta_{eq}^i \hat T A_0
\left[ \frac{\del (f c_1)}{\del \zeta_{eq}^2} \zeta_{eq} \cdot \del \hat T + \frac{\del (f c_2)}{\del \zeta_{eq}^2} \zeta_{eq} \cdot \del \hat \nu +\frac{\del (f c_3)}{\del \zeta_{eq}^2} \zeta_{eq} \cdot \del \zeta_{eq}^2\right]
\end{split}
\end{align}
where all the functions $f,c_i$ are evaluated on $\vec\zeta=\vec\zeta_{eq}$.
We will find Kubo formulas for the $c_i$'s in the next subsection, right after reviewing
the parity odd effective action results of \cite{Bhattacharyya:2012xi}.

\subsubsection{Parity Odd Effective Action}

The most general parity odd (CPT invariant) first order effective action is given by:
\begin{align}
\begin{split} \label{supfl:odd_effective_action}
S^{odd} & = S_1^{odd}+S^{anom} \\
S_1^{odd} & = \int  d^3x \sqrt{g_3} (g_1 \epsilon^{ijk} \zeta_i \del_j A_k +T_0g_2\epsilon^{ijk}\zeta_i\del_j a_k) 
\\
S^{anom} & = \frac{C}{2} \left( \int \frac{A_0}{3T_0} AdA+\frac{A_0^2}{6T_0}Ada \right)
\end{split}
\end{align}
where
\begin{equation}
g_1 = g_1(\hat{T},\hat{\nu},\psi)\ ; \qquad g_2 = g_2(\hat{T},\hat{\nu},\psi)\ ;
\end{equation}
$C$ is the anomaly coefficient and 
\begin{equation}
\hat{\nu}\equiv\frac{\hat{\mu}}{\hat{T}}\ , \qquad \psi \equiv \frac{\zeta^2}{\hat{T}^2}\ .
\end{equation}

The corrections to $J^i$ and $T_0^i$ from the parity odd sector are given by\cite{Bhattacharyya:2012xi}:
\begin{align}
\begin{split}\label{supfl:odd_consist_current_stress}
\delta J^{i} = & \frac{\hat T}{\sqrt{g_3}} \frac{\del S^{odd}}{\del A_i} =   \hat{T} \left(2 g_1 V_6^i +T_0 g_2 V_7^i +\hat{T} g_{1,\hat{T}} V_1^i  - \frac{1}{T_0} g_{1,\hat \nu} V_2^i - g_{1,\psi} V_5^i  \right) \\
& +\frac{2}{\hat{T}} \zeta_{eq}^i (S_1 g_{1,\psi} +T_0 S_2 g_{2,\psi})
+ \frac{C}{3} e^{-\sigma} \left[2 A_0 V_6^i +\frac{A_0^2}{2}V_7^i +\epsilon^{ijk} A_k \del_j A_0 \right],
\\
\delta T_0^i  = & \
 \frac{\hat T}{\sqrt{g_3}} \left[ \frac{\del S^{odd}}{\del a_i\ \  } - A_0 \frac{\del S^{odd}}{\del A_i\ \ } \right]
\\
 = & \ \hat{T} \left( (T_0 g_2 - 2A_0 g_1) V_6^i  - T_0A_0g_2 V_7^i \right)
     -2\frac{A_0}{\hat T} \zeta_{eq}^i (S_1 g_{1,\psi} +T_0 S_2 g_{2,\psi})\\
& + \hat{T}T_0 \left( \hat{T} V_1^i (g_{2,\hat{T}} - \hat{\nu}g_{1,\hat{T}}) - \frac{1}{T_0} V_2^i (g_{2,\hat{\nu}} - \hat{\nu}g_{1,\hat{\nu}})  - V_5^i (g_{2,\psi} - \hat{\nu}g_{1,\psi})\right)  \\
& - \frac{C}{2}A_0^2 e^{-\sigma} \left(V_6^i + \frac{A_0}{3} V_7^i \right) ,
\end{split}
\end{align}
where
\begin{align}
\begin{split}\label{supfl:vis_sis}
S_1  = & \ \epsilon^{ijk} \zeta^{eq}_i \del_j \zeta^{eq}_k, \quad S_2 = \epsilon^{ijk} \zeta^{eq}_i \del_j a_k \\
V_1^i  = & \ \epsilon^{ijk} \zeta^{eq}_j \del_k \sigma, \quad V_2^i = \epsilon^{ijk} \zeta^{eq}_j \del_k A_0,
    \quad V_5^i = \epsilon^{ijk} \zeta^{eq}_j \del_k \psi_{eq},\\
V_6^i  = & \ \epsilon^{ijk}\del_j A_k, \quad V_7^i = \epsilon^{ijk}\del_j a_k\ ,
\end{split}
\end{align}
and all the thermal functions and their derivatives are evaluated at $\vec \zeta= \vec\zeta_{eq}$. A comma followed by a subscript indicates derivative w.r.t the appropriate thermal parameter.
The parity odd one derivative contribution to the covariant current is given by:
\begin{align}
\begin{split}\label{supfl:cov_odd_current}
\delta\tilde{J}^i  = & \ \hat{T} \left( 2 g_1 V_6^i +T_0 g_2 V_7^i +\hat{T} g_{1,\hat{T}} V_1^i - \frac{1}{T_0}g_{1,\hat \nu} V_2^i -g_{1,\psi} V_5^i  \right) \\
& +\frac{2}{\hat{T}} \zeta_{eq}^i (S_1 g_{1,\psi} +T_0 S_2 g_{2,\psi})
+ C e^{-\sigma} \left(A_0 V_6^i +\frac{A_0^2}{2}V_7^i \right).
\end{split}
\end{align}

All the non-dissipative parity odd superfluid transport coefficients can be expressed in terms of the thermal functions $g_1$ and $g_2$. We will find Kubo formulas for those thermal functions in the next subsection.

\subsection{Extracting the Kubo Relations}\label{subsec:supfl:Kubo}
In this subsection we will use our procedure to extract Kubo formulas for the parity even thermal functions $c_1$, $c_2$ and $c_3$.
We will also present Kubo formulas for the parity odd thermal functions $g_1$ and $g_2$.

\subsubsection{Kubo Formulas for the Parity Even Thermal Functions}

Due to the addition of non-local terms to the Goldstone solution it turns out that in the parity even sector one should adopt a slightly different approach to derive Kubo formulas.
It is possible to express the Kubo formulas in terms of correlation functions of the Goldstone phase gradient and another (composite) operator by varying \eqref{supfl:solution_eom} according to \eqref{anom:JT_formulas}.

We get the following Kubo formulas:
\begin{equation}
\begin{split} \label{supfl:c-is}
& \boxed{c_1  = -\frac{1}{T^2} \lim_{\vec k ,\omega \rightarrow 0}  \langle \phi(\vec k) \ T^{00}(-\vec k) \rangle_{\vec\zeta_0\perp\vec k}}
\\
& \boxed{c_2 = -\lim_{\vec k ,\omega \rightarrow 0}  \langle \phi(\vec k) \ J^{0}(-\vec k) \rangle_{\vec\zeta_0\perp\vec k}}
\\
&\boxed{ c_3 = \frac{i}{2T\zeta_0^y} \lim_{\vec{k}\rightarrow 0 }\frac{\del}{\del k_x}\langle \del_x\phi (k_x) \ J^{y} (-k_x);
\vec\zeta \parallel \hat y \rangle}\ ,
\end{split}
\end{equation}
where every correlation function has to be calculated at $\vec \zeta_0\perp \vec k$ (the zero momentum limit should be taken at the last step).

It is also useful to get a Kubo formula for the zeroth order thermal function $f$ from a variation of the zeroth order current.
Let us start by varying $J^j$ with respect to $A_i$. After setting the external sources to their flat space constant values ($T_0,\mu_0,\zeta^i_0$), we get in momentum space for $\vec\zeta_0\perp\vec k$:
\begin{equation}
\begin{split}
G^{i,j}_\perp (k,-k) \biggr{|}_{\omega=0} &\ = \int d^3 x\  e^{-ikx}\left[  \frac{\delta J^j (0)}{\delta A_i(x)} \right] =
- \int d^3 x\  e^{-ikx}\left[ \frac{\delta (f\zeta^j_{eq})}{\delta \zeta^k_{eq}} \frac{\delta \zeta^k_{eq}}{\delta A_i}\right] \\
&\ = -  \left[\frac{\del f}{\del \zeta^2} \cdot 2 \zeta_0^j (\zeta_0)_k + f\delta^j_k \right]\left[\delta^{ki} - \frac{k^k k^i}{k^2}\right]
 + O(k)\\
&\ = -  \left[\frac{\del f}{\del \zeta^2} \cdot 2 \zeta_0^i \zeta_0^j + f\delta^{ij} -f \frac{k^i k^j}{k^2}\right]
+\cancel{(\zeta_0 \cdot k) (\dots)} + O(k)
\ ,
\end{split}
\end{equation}
where we have only used the lowest order solution for the gradient of the Goldstone phase.
If we now take $i=j=x$, $\vec k$ in the $\hat z$ direction and $\vec\zeta_0$ in the $\hat y$ direction (when evaluating correlators in terms of Feynman diagrams, this is our choice to make), we end up with the following Kubo formula for $f$:
\begin{equation}\label{supfl:f}
\begin{split}
\boxed{ f = - 2 \frac{\del P}{\del \zeta^2} = - \lim_{\vec{k},w \rightarrow 0} G^{x,x} (\vec k_z,-\vec k_z; \vec \zeta_0 \parallel \hat y) } \ .
\end{split}
\end{equation}
It is essential that the zero momentum limit is taken after evaluating the formula with $\vec \zeta_0 \perp \vec k$.

One may wonder about the consistency of the derivative expansion when considering non-local terms (of negative momentum powers). Fortunately, if we take the momenta in the direction of one of the axes only, and since $\zeta_{eq}^i$ starts at zeroth order in momenta \eqref{supfl:perp_sol_eom}, we can still count powers of derivatives in a consistent way.\footnote{Additional bookkeeping is required since $\zeta_{eq}^i$ is not of unique order in the derivative expansion.}

It is understood that our Green functions are evaluated in flat space with compactified time coordinate. We can therefore lose the $0$ subscripts on $T,\mu,\vec \zeta$ and present the Kubo formulas as in the introduction (eq.\eqref{intro:even_kubo}).

It is possible to use a similar calculation to obtain Kubo formulas for all the zeroth order thermal functions (energy density, pressure, charge density, entropy density, charge susceptibility, etc.) in all the cases studied in this paper.

Note, that the parity even Kubo formulas derived in this section received no contributions from the parity odd sector. Similarly, the parity odd Kubo formulas that will be derived in the next subsection will receive no parity even contributions. We could have therefore treated the two sectors separately. 

We will give some details on how the hydrodynamic transport coefficients relate to $c_1$, $c_2$ and $c_3$ in the next subsection, right after extracting Kubo formulas for the thermal function $g_1$, $g_2$ of the parity odd sector.

\subsubsection{Kubo Formulas for the Parity Odd Thermal Functions}\label{subsec:supfl:p_odd_Kubo}
In this subsection we obtain Kubo formulas for the thermal functions $g_1$ and $g_2$ from the parity odd effective action of equation \eqref{supfl:odd_effective_action}.
For this purpose we will vary the covariant current (\ref{supfl:even_1_current}, \ref{supfl:cov_odd_current}), and stress tensor (\ref{supfl:even_1_stress}, \ref{supfl:odd_consist_current_stress}) with respect to the gauge field $A_i$ and metric perturbation $a_i$.
In appendix \ref{app:solving_eom} equations (\ref{app:eq:zeta_momentum}, \ref{app:eq:full_phi_sol}, \ref{app:eq:perp_phi_sol}, \ref{app:eq:paral_phi_sol}) we have solved for $\vec\zeta_{eq}$ up to first order in variation of the background fields including the non-local contributions (of negative derivative powers). It will be useful in the following to have expressions for the variation of the superfluid velocity w.r.t to $A_i$ and $a_i$. In the special case of $\vec \zeta_0 \perp \vec k$ we have after setting the source fields to zero:\footnote{Note that here $\zeta_i$ is associated with a momenta $-k$ as in our Kubo formulas derivation of section \ref{sec:anom}, this is the origin of the extra minus sign in \eqref{supfl:del_zeta_del_A}.}
\begin{equation}\label{supfl:del_zeta_del_A}
\frac{\delta\zeta^i_{eq}}{\delta A_j} = \delta^{ij}-\frac{k^ik^j}{k^2} - 2iT_0c_3 k^i \zeta_0^j + O(k^2)\ ,
\end{equation}
and for $\vec \zeta_0 \parallel \vec k$:
\begin{equation}
\frac{\delta\zeta^i_{eq}}{\delta A_j} = \delta^{ij}-\frac{k^ik^j}{k^2}  + O(k^2)\ .
\end{equation}
In both cases we have $\delta\zeta^i_{eq}/ \delta a_j = 0$ at first order in momenta.
In the absence of sources we set $\vec\zeta_{eq} =\vec\zeta_{0}$.

Let us start by varying the covariant current with respect to $A_j$. We get in momentum space, after setting the external sources to zero:
\begin{equation}
\begin{split}
\frac{\delta \tilde J^i}{\delta A_j}=& 2(\zeta_0)_k \frac{\delta \zeta_{eq}^k}{\delta A_j}
\left(- \zeta_0^i \frac{\delta f}{\delta \zeta^2} \left[1 -2iT_0c_3 (\zeta_0 \cdot k) \right]
+fT_0c_3ik^i +\frac{ig_{1,\psi} }{T_0}\epsilon^{ilm} (\zeta_0)_l k_m \right)
\\&
-f\frac{\delta \zeta_{eq}^i}{\delta A_j} + \epsilon^{ijk} ik_k (2g_1 T_0 +C \mu_0)
-\frac{2i}{T_0}\zeta_0^i g_{1,\psi} \epsilon^{mnj} \zeta_m k_n + O(k^2)
,
\end{split}
\end{equation}
where all the functions and their derivatives are evaluated in terms of the flat space parameters $(T_0,\nu_0,\zeta_0^2)$.
This evaluates to:
\begin{equation}
\begin{split}
\frac{\delta \tilde J^i}{\delta A_j}=&
-f(\delta^{ij}-\frac{k^ik^j}{k^2})
-2 \frac{\delta f}{\delta \zeta^2} \zeta_0^j \zeta_0^i
+ \epsilon^{ijk} ik_k (2g_1 T_0 +C \mu_0)
\\&
-4 i
\frac{g_{1,\psi} }{T_0}(\zeta_0)^{[i}\epsilon^{j]mn} (\zeta_0)_m k_n
+ O(k^2)\ ,
\end{split}
\end{equation}
for $\vec \zeta_0 \perp \vec k$, and to:
\begin{equation}
\begin{split}
\frac{\delta \tilde J^i}{\delta A_j}=&
-f(\delta^{ij}-\frac{k^ik^j}{k^2})+ \epsilon^{ijk} ik_k (2g_1 T_0 +C \mu_0)+ O(k^2)\ ,
\end{split}
\end{equation}
for $\vec \zeta_0 \parallel \vec k$. It should be noted that the derivatives of the thermal variables $T_0$, $\mu_0$ and $\zeta^i_0$ vanish at thermal equilibrium in flat space, although the functional derivatives may be non-zero.
Note that these expressions are symmetric under $i\leftrightarrow j$ and $k\leftrightarrow -k$ as they should.
Relating to the current-current Green function using \eqref{anom:JT_formulas} and contracting with the Levi-Civita symbol we get for $\vec \zeta_0 \perp \vec k$:
\begin{equation}
\begin{split}
\epsilon_{ijn}\widetilde G^{j,i}_{\perp} (k_n,-k_n) \biggr{|}_{\omega=0} = 2 ik_n (2g_1 T_0 +C \mu_0)
-4 i \epsilon_{ijn}
\frac{g_{1,\psi} }{T_0}(\zeta_0)^{[i}\epsilon^{j]mn} (\zeta_0)_m k_n
+ O(k^2)\ ,
\end{split}
\end{equation}
and for $\vec \zeta_0 \parallel \vec k$:
\begin{equation}
\begin{split}
\epsilon_{ijn}\widetilde G^{j,i}_{\parallel} (k_n,-k_n) \biggr{|}_{\omega=0} =
2 ik_n (2g_1 T_0 +C \mu_0)+ O(k^2)\ ,
\end{split}
\end{equation}
where the $\perp / \parallel$ subscripts are there to remind us that the Green functions are to be evaluated with superfluid velocity thermal parameter $\vec\zeta_0$ perpendicular/parallel to the external momentum $\vec k$. Setting $\omega=0$ allows us to disregard any dissipative contribution that may arise.

We can now find Kubo formulas using both the perpendicular and the parallel Green functions.
Pursuing both ways will lead us to a new type of identities.
First let us pick $\vec{k} \parallel \vec{\zeta}$. Dividing by $k_n$ and taking the zero momentum limit we get:
\begin{equation} \label{supfl:T2}
\boxed{ C \mu_0+2T_0 g_1=\lim_{k_n\rightarrow0}\frac{i}{2k_n}\epsilon_{ijn} \widetilde G_{\parallel}^{i,j} (k_n,-k_n) \biggr{|}_{\omega=0} }\ .
\end{equation}

The expression we get for $g_1$ is therefore:
\begin{equation}\label{supfl:g1}
\boxed{
g_1=-\frac{C}{2} \left( \frac{\mu_0}{T_0} \right)
+\lim_{k_n\rightarrow0}\frac{i}{4 T_0 k_n}\epsilon_{ijn} \widetilde G_{\parallel}^{i,j} (k_n,-k_n) \biggr{|}_{\omega=0}}\ .
\end{equation}

Had we chosen $\vec{k}\perp \vec\zeta_0$ we would have gotten:
\begin{equation}
\begin{split}
 \frac{\del}{\del \psi_0}\left( g_1 \psi_0  \right) = -\frac{C}{2} \left( \frac{\mu_0}{T_0} \right)+  \lim_{k_n\rightarrow 0} \frac{i}{4 T_0k_n} \sum_{ij} \epsilon_{ijn}\widetilde G_{\perp}^{i,j} (k_n,-k_n) \biggr|_{\omega=0}
    \ ,
\end{split}
\end{equation}
where $\psi_0\equiv\frac{\zeta_0^2}{T_0^2}$.
There is a slight abuse of notation in the last formula (and similar formulas above) in the sense that it is not clear what exactly we mean by the $k_n$ division in the last equation. What we mean is that the momentum in the Green function should be taken in the $n$ direction (which is our choice to make), we then divide by the same $k_n$ and take the zero momentum limit. An explicit calculation could use for example $G^{x,y}$ with $\vec k$ in the $\hat z$ direction ($n=z$), and with perpendicular $\vec \zeta_0$ in the $\hat x$ or $\hat y$ directions. No summation over $n$ is implied, but we could have used a very similar formula with summation over $n$.
The $\psi$ differentiation was taken at constant $T$ and $\nu$, so integrating back, and losing all the $0$ subscripts everywhere, we get
\begin{equation}
\begin{split}
\boxed{
g_1 = -\frac{C}{2} \left( \frac{\mu}{T} \right) + \frac{1}{\psi}
 \int d\psi \left[ \lim_{k_n\rightarrow 0} \frac{i}{4Tk_n} \sum_{ij} \epsilon_{ijn}  \widetilde G_{\perp}^{i,j} (k_n,-k_n) \right]
+\frac{F(T,\nu)}{\psi}
    }\ ,
\end{split}
\end{equation}
where $F(T,\nu)$ could be any arbitrary function of $T$ and $\nu$ and the correlator is evaluated in flat space with temperature $T$, chemical potential $\mu$ and transverse superfluid velocity $\zeta^i$. This is not a full determination of $g_1$, but we nevertheless find it interesting because of the identity that follows from it.

Comparing this to the last formula we got for $g_1$ we reach the conclusion that
\begin{equation}
\begin{split}
\boxed{
 \lim_{k_n\rightarrow 0}  \frac{i}{k_n} \epsilon_{ijn}
\left [ -  \psi \widetilde G_{\parallel}^{i,j} (k_n,-k_n)  +\int d\psi \widetilde G_{\perp}^{i,j} (k_n,-k_n) \right]
 \biggr{|}_{\omega=0}=
-4 T F(T,\nu) }\ .
\end{split}
\end{equation}
The fact that this combination of Green functions does not depend on the superfluid transverse velocity component $\zeta^2$ is curious and deserves further study.

Let us now proceed to obtain the Kubo formula for the thermal function $g_2$. We will keep using the parallel limit $\vec \zeta_0 \parallel k$ which leads to simpler Kubo formulas. Looking at the $\widetilde G^{0i,j} = - \left( ( {\delta \tilde J^j}/{\delta a_i} ) - A_0 ( {\delta \tilde J^j} / {\delta A_i} ) \right)$ correlator (evaluated in flat space) we get:
\begin{equation}\label{supfl:T3}
\boxed{ C\mu_0^2+4 T_0 \mu_0 g_1 - 2T_0^2 g_2 =  \lim_{k_n \rightarrow 0} \frac{i}{k_n} \epsilon_{ijn} \widetilde G^{0i,j}_\parallel (k_n,-k_n) }\ .
\end{equation}

Using the expression we already found for $g_1$ we get:
\begin{equation}\label{supfl:g2}
\boxed{
g_2 =
-\frac{C}{2} \left( \frac{\mu_0}{T_0} \right)^2 -  \epsilon_{ijn}
\lim_{k_n \rightarrow 0} \frac{i}{2T_0^2 k_n} \left [    \widetilde G^{0i,j}_\parallel (k_n,-k_n)
-{\mu_0} \widetilde G_{\parallel}^{i,j} (k_n,-k_n) \right] \biggr{|}_{\omega=0} }\ .
\end{equation}

For the clarity of structural arguments that we intend to make later, let us consider what would change in our analysis when including the CPT violating term \cite{Bhattacharyya:2012xi}
\begin{equation}
\delta S^{odd}  = C_1 T_0^2 \int d^3 x \sqrt{g_3} \epsilon^{ijk} a_i \del_j a_k\ ,
\end{equation}
in the parity odd superfluid effective action \eqref{supfl:odd_effective_action}. When $C_1$ is a dimensionless constant this term respects all the required symmetries except CPT.
Such a term would not change the charge current, and would have therefore no effect on the Kubo formulas derived above.
It would, however, change the stress-energy tensor $T^i_0$ by the additional term:
\begin{equation}
\delta T^{i}_{0} = 2 C_1 e^{-\sigma} T_0^3 V_7^i\ .
\end{equation}
The $\epsilon_{ijn} G^{0i,0j}$ correlator will allow us to derive a Kubo formula for $C_1$.
Since $T^{0j} = (T_0^j - g_{0k} T^{jk})/g_{00}$ and since $T^{jk}$ is symmetric and  $g_{0k}$ vanishes in the absence of sources, no contribution to the correlator comes from the second term. We are therefore left with:
\begin{equation}
\epsilon_{ijn} G^{0i,0j} = \epsilon_{ijn} \left(\frac{\delta}{\delta a_i}-A_0\frac{\delta}{\delta A_i}\right) T_0^j\ .
\end{equation}
Evaluating the expression with $\vec{k}\parallel \vec \zeta_0$ we get:
\begin{equation} \label{supfl:T4}
\boxed{ \frac{1}{3}C\mu_0^3 -2 T_0^2 \mu_0 g_2 + 2T_0 \mu_0^2 g_1 +2C_1 T_0^3 = \lim_{k_n \rightarrow 0} \frac{i}{2k_n} \epsilon_{ijn} G^{0i,0j}_\parallel (k_n,-k_n)}\ .
\end{equation}
Isolating $C_1$:
\begin{equation}\label{supfl:bigC_1}
\boxed{
C_1 = -\frac{C}{6}\left(\frac{\mu_0}{T_0}\right)^3
+ \lim_{k_n \rightarrow 0} \frac{i}{4T_0^3k_n} \epsilon_{ijn}
\left[ G^{0i,0j}_\parallel
- 2 \mu_0 \widetilde G_{\parallel}^{0i,j}
+ \mu_0^2 \widetilde G^{i,j}_\parallel \right] \biggr{|}_{\omega=0} }\ ,
\end{equation}
where all correlators are evaluated with $(k_n,-k_n)$ external momenta.

\subsubsection{Generalization to Multiple Superfluid Charges} \label{subsubsec:supfl:MultipleCharges}
A generalization of our analysis to superfluids with multiple unbroken charges (but only one broken charge) seems straightforward at least in the parity odd sector. The same case was treated in \cite{Neiman:2011mj}. For simplicity we will only be considering multiple Abelian charges (a tensor product of multiple $U(1)$-s). A
non-Abelian generalization is very likely possible.
Our goal in this subsection is to reveal the charge-index structure of the formulas we have presented in the previous subsection. This by no means constitutes a full treatment of superfluids with multiple broken charges.

First, we have to replace the first order parity odd effective action with a multiple-charge extension of the form:
\begin{align}
\begin{split}\label{supfl:multiple_podd_action}
S^{odd} & = S_1^{odd}+S^{anom} \ ,\\
S_1^{odd} & = \int  d^3x \sqrt{g_3} (g^{ab}_1 \epsilon^{ijk} \zeta_i^a \del_j A_k^b +T_0g_2^a \epsilon^{ijk}\zeta_i^a\del_j a_k) +\frac{C_1 T_0^2}{2} \int{ada}\ ,
\\
S^{anom} & = \frac{C^{abc}}{2} \left( \int \frac{A_0^a}{3 T_0} A^b dA^c +\frac{A^a_0 A_0^b}{6 T_0}A^cda \right)\ ,
\end{split}
\end{align}
where $a,b,c$ are charge indexes. The index associated with the broken charge is $a=0$. Only one superfluid transverse velocity exist which is associated with the broken charge $\zeta_i^{a=0}$. All normal charges are related to appropriate gauge covectors $\zeta_i^{a\neq0} = A^a_i$. $g_1^{ab}$ should vanish for $a\neq 0$, and $g_2^a$ should become a constant in that case.
The requirement of CPT invariance of the partition function forces $C_1 = 0$.
This would result in the following Kubo formulas for $g_1^{(ab)}$, $g_2^a$ and $C_1$:
\begin{align}
&\boxed{
g_1^{(ab)}= -\frac{C^{abc}}{2} \left( \frac{\mu_0^c}{T_0} \right)
+\frac{i}{4 T_0}  \lim_{k_n\rightarrow0} \epsilon_{ijn} \del_{k_n}  \widetilde G_{\parallel}^{ai,bj} (k_n,-k_n) \biggr{|}_{\omega=0}}\ , \label{supfl:kubo_multip_g1}
\\
&\boxed{
g_2^a =
-\frac{C^{abc}\mu_0^b \mu_0^c}{2 T_0^2} -  \frac{i}{2T_0^2}
\lim_{k_n \rightarrow 0}\epsilon_{ijn} \del_{k_n} \left [    \widetilde G^{0i,aj}_\parallel (k_n,-k_n)
-{\mu_0^b} \widetilde G_{\parallel}^{ai,bj} (k_n,-k_n) \right] \biggr{|}_{\omega=0} }\ ,
\\
&\boxed{
C_1 = -\frac{C^{abc}\mu_0^a\mu_0^b\mu_0^c}{6T_0^3}
+ \frac{i}{4T_0^3} \lim_{k_n \rightarrow 0} \epsilon_{ijn} \del_{k_n}
\left[ G^{0i,0j}_\parallel
- 2 \mu^a_0 \widetilde G_{\parallel}^{0i,aj}
+ \mu_0^a \mu_0^b \widetilde G^{ai,bj}_\parallel \right] \biggr{|}_{\omega=0} }\ ,
\end{align}
where $G^{ai,bj}$ and $G^{0i,aj}$ are defined in a similar way to the one described in \eqref{anom:Green_def}, adding the appropriate charge indexes on the $\mathcal A_\mu$ derivatives. $C^{abc}$ is the completely symmetric anomaly coefficient of three currents. $g_1^{(ab)}$ is the symmetric part of $g_1^{ab}$.\footnote{This however constitutes a full determination of $g_1^{ab}$ since only $g_1^{0b}$ can be non-vanishing.}
We have replaced the $k_n$ division of equations (\ref{supfl:g1}, \ref{supfl:g2}, \ref{supfl:bigC_1}) by a $\del_{k_n}$ differentiation in the above formulas. We find this form more likely to be generalized to the case of superfluid with multiple broken charges since the differentiation makes sure that we get rid of any zeroth order contribution that may arise.
The above Kubo formulas reveal the full charge-index structure of the formulas derived in the last subsection.

In the case of more than one broken charge a bunch of new scalars are available at zeroth order for constructing the effective action due to mixed products of different-charge superfluid transverse velocities of the form $\zeta_a \cdot \zeta^b$. Therefore a generalized new analysis is needed, even at zeroth order, to constitute a full treatment of superfluids with multiple broken charges.
It is important to emphasize that we have not listed all the possible contribution to the effective action of a superfluid with multiple broken charges in equation \eqref{supfl:multiple_podd_action}, even in the parity odd sector ($\int d^3x\sqrt{g_3}\kappa^{abc}_1 \epsilon^{ijk} \zeta_i^a \zeta_j^b \zeta_k^c$, $\int  d^3x \sqrt{g_3} \kappa^{ab}_2 \epsilon^{ijk} \zeta_i^a \zeta_j^b \del_k \hat T$ were ignored, just to name a few).
In addition, for the case of multiple broken charges the full relation between $g_1^{ab}$, $g_2^a$ and $C_1$ and the (non-dissipative) hydrodynamic superfluid transport coefficients hasn't been studied yet. For a non-Abelian analysis one has to furthermore extend the derivatives to covariant derivatives and check the influence of this change.
We leave this for a future study.

\subsection{Hydrodynamic Transport Coefficients} \label{subsec:supfl:transports}
In this subsection we present the relations between the thermal functions $c_1,c_2,c_3$, $g_1,g_2$ and the non-dissipative part of the superfluid constitutive relations. The Kubo formulas for the thermal functions were already found in the previous subsection. Having this in hand, and stating the constitutive relations, we can identify Kubo formulas for any of the superfluid non-dissipative transport coefficients.

\subsubsection{Parity Even Transport Coefficients}
We start with the parity even sector.
The parity even superfluid constitutive relations are the first order parity even corrections to stress tensor $\pi^{\mu\nu}$, charge current $j^\mu_{diss}$ and ``Josephson equation'' $\mu_{diss}$. The expressions are given in terms of the hydrodynamic fields $T,\mu,\zeta^\mu,u^\mu$ and derivatives thereof.

The superfluid constitutive relation we shall present are given in terms of some special combinations of $\pi^{\mu\nu}$, $j^\mu_{diss}$, $\mu_{diss}$ that are invariant under frame redefinitions (this is sometimes more convenient as was explained at the end of subsection \ref{subsec:supfl:prelimI}). To completely specify the constitutive relations one has to specify five additional frame fixing conditions.
Transforming between two fluid frames is a simple task (see section (2.4) of \cite{Bhattacharya:2011eea} for a detailed discussion).

The constitutive relations are expressed in terms of the thermal functions $c_1,c_2,c_3$. Since the Kubo formulas for these thermal functions were already found \eqref{supfl:c-is}, we now have in hand Kubo formulas for all the parity even non-dissipative superfluid transport coefficients.
The results for the constitutive relations are taken from \cite{Bhattacharyya:2012xi}.

The frame redefinition invariant combinations that are used to present the constitutive relations are:
\begin{align}
\begin{split}
{\mathcal S}_a =&\  \left(\frac{s}{\epsilon + P}\right)\frac{\partial}{\partial H_a}\left(\frac{q}{s}\right) \left[-\left(\frac{u_\nu \zeta_\mu \pi^{\mu\nu}}{T}\right)+ \nu \left(\pi^{\mu\nu}P_{\mu\nu} - \frac{3}{2}\pi^{\mu\nu}\tilde P_{\mu\nu} \right) + \frac{\epsilon+P}{T}  \mu_{diss}\right] \\
&\ +\left(\frac{1}{s}\frac{\partial s}{\partial H_a}\right)\left( \frac{\pi^{\mu\nu}\tilde P_{\mu\nu}}{2T}\right)-\left(\frac{u_\mu u_\nu \pi^{\mu\nu}}{T^2}\right)\delta_{a,1} +(j_{diss}\cdot u)\delta_{a,2}\\
&\ + \left(\frac{1}{2 T \zeta^2}\right) \left(\pi^{\mu\nu}P_{\mu\nu} - \frac{3}{2}\pi^{\mu\nu}\tilde P_{\mu\nu}\right)
  \delta_{a,3}
\end{split}\\
\begin{split} \notag
{\mathcal S}_4= & \ j_{diss}\cdot\zeta + R u_\mu \zeta_\nu \pi^{\mu\nu} + (1-\mu R)\left(\pi^{\mu\nu}P_{\mu\nu} - \frac{3}{2}\pi^{\mu\nu}\tilde P_{\mu\nu}\right)\\
{\mathcal V}_{1\mu} =&\ \left(j_{diss}^\nu + R u_\alpha\pi^{\alpha\nu} \right) \tilde P_{\nu\mu}\\
{\mathcal V}_{2\mu} =&\ \zeta_\alpha \pi^{\alpha\nu} \tilde P_{\nu\mu}\\
\mathcal{T}^{\mu\nu} =&\  \tilde P^{\mu\alpha}\tilde P^{\nu\beta}\left[\pi_{\alpha\beta} -\frac{\eta_{\alpha\beta}}{2}\left(\tilde P_{\theta\phi}\pi^{\theta\phi}\right)\right]\ ,
\end{split}
\end{align}
where $a=\{1,2,3\}$ and:
\begin{equation}
\begin{split}
& H_1 = T; \quad H_2 = \nu ; \quad H_3=  \zeta^2 ; \\
& R=\frac{q}{\epsilon+P}; \quad \tilde P^{\mu\nu} = P^{\mu\nu} - \frac{\zeta^\mu \zeta^\nu}{\zeta^2} .
\end{split}
\end{equation}
A minor typo in $\mathcal{S}_a$ (minus sign in the first term) was corrected here (compared to \cite{Bhattacharyya:2012xi}).

Using these, the constitutive relations are (we only present the non-dissipative part which is fixed by the equilibrium partition function):
\begin{align}
\begin{split}
{\mathcal S}_a =& \ -\sum_{b=1}^3 (\zeta \cdot \partial H_b)\bigg\{f\left(\frac{\partial c_b}{\partial H_a} - \frac{\partial c_a}{\partial H_b}\right)-\frac{f c_a}{T} \delta_{b,1} + \frac{f c_b}{\zeta^2}\delta_{a,3}\\
& \qquad +c_b \left[s \frac{\partial}{\partial H_a}\left(\frac{f}{s}\right)+\left( \frac{fT s\nu}{\epsilon +P}\right) \frac{\partial}{\partial H_a}\left(\frac{q}{s}\right) \right]\bigg\}+ diss\ ,
\\
{\mathcal S}_4 =& -\sum_b(\zeta\cdot\partial H_b)  f T (1-\mu R)c_b + diss\ ,\\
{\mathcal V}_{1\mu}=& T(1-\mu R)f\sum_b  c_b \tilde P_\mu^\nu\partial_\nu H_b+ diss\ ,\\
{\mathcal V}_{2\mu}=&\ -T\zeta^2f\sum_b  c_b \tilde P_\mu^\nu\partial_\nu H_b+ diss\ ,\\
\mathcal{T}^{\mu\nu} =&\ diss\ ,
\end{split}
\end{align}
where $diss$ stands for additional dissipative terms.

A minor typo of \cite{Bhattacharyya:2012xi} was corrected here by an additional $fT$ factor in the last term of the constitutive relations for ${\mathcal S}_a$.

Using this in the transverse frame one obtains the following expression for the current:
\begin{align}
\begin{split}\label{supfl:j_diss_even_landau}
(j_{diss})_\mu = &
\frac{T^2sf}{\epsilon+P}P_\mu^\nu (c_a\del_\nu H_a)
+ \frac{2T^2s}{\epsilon+P}  \zeta_\mu
\left[ \zeta \cdot \del H_a \frac{\del(fc_a)}{\del\zeta^2} \biggr|_{s,\frac{q}{s}}
 - \frac{f}{T}\zeta \cdot \del(Tc_3) \right.
 \\
 &\left.
+\frac{f}{T}\zeta \cdot \del (Tc_1)
        \left[ \frac{\frac{\del s}{\del \zeta^2} \frac{\del\frac{q}{s}}{\del \nu}-\frac{\del s}{\del \nu} \frac{\del\frac{q}{s}}{\del \zeta^2} }
                    {\frac{\del s}{\del T} \frac{\del\frac{q}{s}}{\del \nu} -\frac{\del s}{\del \nu} \frac{\del\frac{q}{s}}{\del T}} \right]
+\frac{f}{T}\zeta \cdot \del (Tc_2)
        \left[ \frac{ \frac{\del s}{\del T} \frac{\del \frac{q}{s}}{\del \zeta^2} - \frac{\del s}{\del \zeta^2} \frac{\del\frac{q}{s}}{\del T} }
                    {\frac{\del s}{\del T} \frac{\del\frac{q}{s}}{\del \nu} -\frac{\del s}{\del \nu} \frac{\del\frac{q}{s}}{\del T} } \right] \right]
\\
 =
\frac{T^2s}{\epsilon+P} & \left[f c_a P_\mu^\nu \del_\nu H_a
+ 2  \zeta_\mu \zeta^\nu \left[
  \del_\nu H_a \frac{\del(fc_a)}{\del\zeta^2} \biggr|_{s,\frac{q}{s}}
 - \frac{f}{T} \frac{\det{\left(\frac{\del s}{\del H_a}; \frac{\del \frac{q}{s} }{\del H_a} ; \del_\nu(Tc_a)\right)}}
 {\det{\left(\frac{\del s}{\del H_a}; \frac{\del \frac{q}{s} }{\del H_a} ; \frac{\del(\zeta^2 )}{\del H_a}\right)}} \right]\right]
\end{split}
\end{align}
where the derivative with explicit subscripts $s$ and $q/s$ is taken as constant $s$ and $q/s$ (in \cite{Neiman:2011mj} it was suggested that the set of variables $(s, q/s, \zeta^2)$ are better suited to describe some properties of superfluid hydrodynamics than $(T, \nu, \zeta^2)$). A summation over a is implied.

\subsubsection{Parity Odd Transport Coefficients}
We now move to the parity odd part of the first order superfluid constitutive relations.
We present them in the transverse frame of \eqref{supfl:frame_fix}.
We find it easier to identify the physical significance of each transport term this way.
The constitutive relations are given in terms of the thermal functions $g_1$, $g_2$, $C_1$ by the following formulas:
\begin{align}
 \begin{split}
   \pi^{\mu\nu} ={}& -sTP^{\mu\nu}\left(2T\zeta_\rho\omega^\rho\left(\frac{\mu}{T}\ddsimple{g_1}{s} - \ddsimple{g_2}{s}\right)
     + \zeta_\rho B^\rho \ddsimple{g_1}{s}\right) \\
    &- 2T\zeta^\mu\zeta^\nu\left(2T\zeta_\rho\omega^\rho\left(\frac{\mu}{T}\ddsimple{g_1}{\zeta^2} - \ddsimple{g_2}{\zeta^2}\right)
     + \zeta_\rho B^\rho \ddsimple{g_1}{\zeta^2} \right) +diss \ ,
  \end{split} \\
 \begin{split}\label{supfl:j_diss_odd_landau}
   j_{diss}^{\mu} ={}& \omega^\mu\left(\vphantom{\frac{n^a}{\epsilon+P}} C\mu^2 + 4 g_1 \mu T - 2 g_2 T^2
       - \frac{2q}{\epsilon+P}\left(\frac{1}{3}C\mu^3 + 2g_1 \mu^2 T - 2 g_2 \mu T^2 + 2 C_1 T^3\right) \right) \\
      &+ B^\mu\left(C \mu + 2Tg_1
          - \frac{q}{\epsilon+P}\left(\frac{1}{2}C \mu^2 + 2 g_1 \mu T -g_2 T^2 \right)\right)
        \\
        & +\frac{2T^2 s}{\epsilon+P} \zeta^\mu
            \left(2T\zeta_\rho\omega^\rho\left(\frac{\mu}{T}\ddsimple{g_1}{\zeta^2} - \ddsimple{g_2}{\zeta^2}\right)
        + \zeta_\rho B^\rho \ddsimple{g_1}{\zeta^2}
        \right)\\
        &
      - \frac{T^2}{\epsilon+P} \epsilon^{\mu\nu\rho\sigma}u_\nu\zeta_\rho\left( q \del_\sigma g_2 + s\del_\sigma g_1 \right)
            +diss \ ,
 \end{split}  \\
 \begin{split}
   \mu_{diss} ={}& \zeta_\mu\omega^\mu\left(\frac{2}{\epsilon+P}\left(\frac{1}{3}C \mu^3 + 2 g_1 \mu^2 T - 2 g_2 \mu T^2 + 2 C_1 T^3 \right) \right. \\
       &\left.{}\ + \frac{4T^2\mu \zeta^2}{\epsilon+P}\left(\frac{\mu}{T}\ddsimple{g_1}{\zeta^2} - \ddsimple{g_2}{\zeta^2} \right)
        - \frac{2T^2}{s}\left(\frac{\mu}{T}\ddsimple{g_1}{(q/s)} - \ddsimple{g_2}{(q/s)}\right) \right) \\
       {}+{}& \zeta_\mu B^\mu\left(\frac{1}{\epsilon+P}\left(\frac{1}{2}C\mu^2 + 2g_1\mu T - g_2 T^2\right)
        + \frac{2T\mu\zeta^2}{\epsilon+P}\ddsimple{g_1}{\zeta^2} - \frac{T}{s}\ddsimple{g_1}{(q/s)} \right) \ .
 \end{split}
\end{align}
These were derived in \cite{Neiman:2011mj}.\footnote{We have noticed a typo in \cite{Neiman:2011mj},  the  $\zeta^\mu$ term of the charge current is missing.} After correcting for this term, the results match precisely those of \cite{Bhattacharya:2011tra}, \cite{Bhattacharyya:2012xi}.
The partial derivatives with respect to $s$, $q/s$ and $\zeta^2$ are taken with $(s, q/s, \zeta^2)$ as the independent thermal parameters. For the full charge-index structure one may refer to \cite{Neiman:2011mj}.
To get these formulas we had to use the following matching rules: $g_1=\alpha=\sigma_8$, $g_2 = -\beta = -\sigma_{10}+2\nu \sigma_8+\frac{1}{2} C \nu^2 +2 \tilde{h}\nu$, $2C_1 = \gamma=s_9$ to match between the different conventions of \cite{Bhattacharyya:2012xi}, \cite{Neiman:2011mj}, \cite{Bhattacharya:2011tra} respectively.

The chiral magnetic and chiral vortical conductivities (i.e. the coefficients of magnetic field and vorticity in the charge current) take the form:
\begin{align}
\begin{split}
\boxed{ \xi_\omega = \vphantom{\frac{n^a}{\epsilon+P}} C\mu^2 + 4 g_1 \mu T - 2 g_2 T^2
       - \frac{2q}{\epsilon+P}\left(\frac{1}{3}C\mu^3 + 2g_1 \mu^2 T - 2 g_2 \mu T^2 + 2 C_1 T^3\right) }\ ,
\end{split}\\
\begin{split}
\boxed{ \xi_B =
C \mu + 2Tg_1
          - \frac{q}{\epsilon+P}\left(\frac{1}{2}C \mu^2 + 2 g_1 \mu T -g_2 T^2 \right)
 }\ ,
\end{split}
\end{align}
and can therefore be expressed (based on our analysis in the previous subsection) using the following Kubo formulas:
\begin{align}
\begin{split}
\boxed{ \xi_\omega =  \lim_{k_n\rightarrow 0} \sum_{ij} \frac{i}{k_n} \epsilon_{ijn} \left[ \widetilde G^{0i,j}_\parallel (k_n,-k_n)  - \frac{q}{\epsilon+P}
 G^{0i,0j}_\parallel (k_n,-k_n) \right] \biggr|_{\omega=0} }\ ,
\end{split}\\
\begin{split}
\boxed{ \xi_B =
 \lim_{k_n\rightarrow 0} \sum_{ij} \frac{i}{2k_n}\epsilon_{ijn}
    \left[ \widetilde G_{\parallel}^{i,j} (k_n,-k_n)   - \frac{q}{\epsilon+P}   \widetilde G^{0i,j}_\parallel (k_n,-k_n)  \right]\biggr|_{\omega=0} }\ .
\end{split}
\end{align}
These formulas strongly resemble the Kubo formula we got for the case of the normal fluid \eqref{anom:vort_kubo}, \eqref{anom:magn_kubo}. The only difference is that the correlators should be evaluated on a thermal background with finite value of the superfluid transverse velocity $\vec \zeta$. As we mentioned earlier, when evaluating a diagram, a finite value of $\vec\zeta$ is expected to influence the propagators as well as the vertices. We will make this statement more precise in the next subsection. The momenta should be taken parallel to $\vec \zeta$.

\section{Discussion}

There are various open issues that deserve further study, and we list some of them below.
It would be interesting to evaluate in field theory models the Kubo formulas that we derived for superfluid transport.
Of particular interest are the chiral magnetic and chiral vortical effects.

Evaluating superfluid Kubo formulas using Feynman diagrams requires the consideration of the new thermal parameter $\zeta_0^i$. The addition of a superfluid velocity strongly resembles the addition of a finite chemical potential to the problem.
Both always appear in the hydrodynamic description accompanied by the appropriate gauge field component (see \eqref{supfl:xi}). This suggests that the new thermal parameter $\vec \zeta_0$ should be introduced to the thermal QFT description the same way that the thermal chemical potential $\mu_0$ is. That is, through an adjusted definition of the grand canonical partition function
obtained from the original partition function by the substitution rule $\mathcal A_\mu \rightarrow \mathcal A_\mu + (\mu_0,\zeta_0^i)$ in the functional integral of the original lagrangian of the theory.

The partition function (and all derived correlation functions) could therefore be calculated using the path integral formalism with time coordinate compactified on a circle of radius $1/T_0$, and where derivatives (momentum vectors in momentum space) are subject to the following substitution rule: $k^\mu \rightarrow (i\omega_n +\mu_0,\vec k + \vec \zeta_0)$, where $\omega_n=\pi T_0 (2n+[1])$ are the bosonic [fermionic] Matsubara frequencies.
Propagators will exhibit a suitable change.

This change is in addition to the usual changes that have to be made when evaluating Feynman diagrams in theories that have a spontaneous symmetry breaking. These include developing the theory in terms of new fields around the vacuum expectation value of the charged scalar operator and using those fields as the new elementary fields of the theory.

One may also wish to evaluate the second order non-conformal normal fluid transport coefficients obtained in appendix \ref{sec:second_order} in the strong coupling limit using AdS/CFT.
It would be interesting to see the effect of these new non-conformal coefficients on observables such as the elliptic flow and multiplicities in numerical hydrodynamic simulations of Heavy-Ion collisions such as \cite{Luzum:2008cw}.

It would be interesting to generalize our results and derive Kubo formulas for the first order non-dissipative transport coefficients of anomalous fluids in arbitrary dimensions using the equilibrium partition function \cite{Banerjee:2012cr}. A similar extension of our analysis will enable the derivation of Kubo formulas for Rindler hydrodynamics at second order using the partition function of \cite{Meyer:2013sva}.
Another required generalization of our work is the derivation of Kubo formulas for superfluids with more than one broken charge.

In \cite{Neiman:2011mj} it was suggested that the hierarchy of charge indexes of the thermal functions/constants $C^{abc}$, $g_1^{ab}$, $g_2^{a}$ and $C_1$
and the associated factors of $\mu$ and $T$ in expressions of the form \eqref{supfl:T2}, \eqref{supfl:T3}, \eqref{supfl:T4},
suggests that our thermal functions/constants may be related to anomaly coefficients of triangular diagrams with the appropriate number of charge current vertices.

The fact that $C^{abc}$ is the anomaly coefficient of the triangular diagram with three currents already came about from entropic constraints (\cite{Son:2009tf,Neiman:2011}). The relation between $g_2^{a}$ and the coefficient of mixed chiral gravitational $JTT$ anomaly was subject to intense debate recently (see e.g. \cite{Landsteiner:2011cp,Landsteiner:2011iq,Chapman:2012my,Golkar:2012kb,Hou:2012xg,Jensen:2012kj}). The relation between $C_1$ and the coefficient of the $TTT$ anomaly is motivated by the fact they both vanish (in the case of $C_1$, due to CPT invariance).

The authors of \cite{Neiman:2011mj} conjectured that in light of the progression of the charge-index structure and the associated factors of $\mu$ and $T$ in the hydrodynamic constitutive relations, $g_1^{ab}$ (their $\alpha^{ab}$) should be related to the coefficient of the $JJT$ anomaly. This led them to conjecture that $g_1^{ab}$ should in fact vanish. This has been proven for the case of a normal fluid (see \cite{Jensen:2012jy}) as it must from $CPT$.\footnote{In \cite{Jensen:2012jy} the normal fluid analog of $g_1^{ab}$ was named $f_1^{AB}$.}

We have tried to repeat the proof of \cite{Jensen:2012jy} for the case of a superfluid.\footnote{In \cite{Jensen:2012jy} the author constrains the structure of the $J^iJ^jJ^0$ three point function using arguments of symmetry and the standard anomalous (non)-conservation equation. The author then relates it to a variation of the $J^iJ^j$ two point function (Kubo formula for the magnetic conductivity) with respect to the chemical potential. Invoking CPT invariance one can then rule out the presence of $\sim T$ term in the magnetic conductivity of a normal fluid.}
Here, due to the possibility of including non local terms (with various powers of momenta in the denominator), we find that it is no longer possible to prove that $g_1^{ab}=0$. One should take into account that the presence of a Goldstone mode allow for long range correlations.
We find that the most general form of the current three-point function is
\begin{equation}
\begin{split}
G^{ai,bj,c0}(k_1,k_2)
 \sim &  -i\epsilon^{ijk} ((k_1)_k \Sigma_1^{0,abc}-(k_2)_k\Sigma_1^{0,bac})
\\& + i \frac{\Sigma_2^{0,abc}\epsilon^{jlk}k_{1l}k_{2k}k_2^i-\Sigma_2^{0,bac}\epsilon^{ilk}k_{1l}k_{2k}k_1^j}{k_1\cdot k_2} + \dots
\end{split}
\end{equation}
where $\Sigma_2^{0,abc}$ encodes $dg_1^{ab}/d\mu^c$.

In general we could use an analysis similar to the one in the previous subsections to relate $JJT$ and $g_1$ motivated by the fact that temperature differentiation is related to $T^{00}$ insertion:\footnote{The second $JJJ$ term in equation \eqref{disc:JJT} is related to the diffeomorphism transformation law of the gauge covector (see last paragraph before eq.(55) in section 4 of \cite{Neiman:2011mj} for more detailed discussion).}
\begin{align}
\begin{split}\label{disc:JJT}
\langle J^i(-\vec k)J^j(\vec k)T^{00}(0) \rangle_{\parallel} =&  T \frac{d}{dT} \langle J^i(-\vec k)J^j(\vec k)\rangle_{\parallel} +
\mu\langle J^i(-\vec k)J^j(\vec k)J^{0}(0) \rangle_{\parallel}  \\& =
\epsilon^{ijk}ik_k (C\mu + 2g_1 T + 2g_{1,T} T^2  - 4g_{1,\psi} \psi  T) + {\mbox{p. even} \atop \mbox{terms}}
\end{split}
\end{align}
We therefore cannot find a general reason why $g_1$ should vanish in theories with non-finite correlation length. We can however generally relate it to the $JJT$ diagram as suggested by \cite{Neiman:2011mj}.

Finally, it would also be interesting to derive the Kubo formulas for the dissipative hydrodynamic coefficients.
This requires to study time dependent dynamics as was done in \cite{Moore:2010bu}. One can repeat our analysis of the parity odd sector omitting the non-local terms, and get precisely the same Kubo formulas as we got in subsection \ref{subsec:supfl:p_odd_Kubo}.
Drawing the conclusions from this, it is possible that the Goldstone phase gradient could be treated as an independent parameter without having to solve for it in terms of the external sources in the parity odd sector even in the dissipative case. All this is true up to an arbitrary addition that vanishes using the Goldstone equation of motion. This might facilitate the Kubo derivation for the dissipative superfluid transport coefficients in the parity odd sector. One such transport coefficient of special interest is the chiral electric conductivity of \cite{Neiman:2011mj}.

\section*{Acknowledgements}
The authors would like to thank Sayantani Bhattacharya, Yasha Neiman, Giuseppe Policastro, and Liran Rotem for valuable discussions and comments. This work was supported in part by the ISF Center of Excellence, the I-CORE program of Planning and Budgeting Committee, and the Ministry of Science and Technology, Israel.

\appendix

\section{Goldstone Equation of Motion at Next to Leading Order in Derivatives}\label{app:Goldstone_eom_corrections}
The full Goldstone effective action up to first order in derivatives can be written as:
\begin{equation}
\begin{split}\label{app:eq:Goldsteom:part_supfl_full}
S &\ = S_0 + S_1^{even}+S_1^{odd}+S^{anom}\ ,\\
S_0&\ = \int d^3 x \sqrt{g_3} \frac{1}{\hat T} P(\hat T, \hat \mu , \xi^2)\ ,\\
S_1^{even}&\  = \int d^3 x \sqrt{g_3} \left[ f c_1 (\zeta \cdot \del) \hat T
+ f c_2 (\zeta \cdot \del) \hat \nu
+ f c_3 (\zeta \cdot \del) \zeta^2   \right]\ ,\\
S_1^{odd} & = \int  d^3x \sqrt{g_3} (g_1 \epsilon^{ijk} \zeta_i \del_j A_k +T_0g_2\epsilon^{ijk}\zeta_i\del_j a_k) \ ,
\end{split}
\end{equation}
where $\zeta_i = A_i-\del_i \phi$. To get the Goldstone equation of motion we have to vary with respect to the Goldstone phase $\phi$. Since $S^{anom}$ does not depend on $\phi$, we will not need its explicit form. The equation of motion up to this order in derivatives reads:
\begin{equation}
\begin{split}\label{app:eq:Goldsteom:eom_supf_full}
0 = \nabla_i  & \left( -\frac{f}{\hat T}\zeta^i +
2\zeta^i \frac{\del(fc_a)}{\del\zeta^2} \zeta \cdot \del H_a
+ fc_a \del^i H_a - 2\zeta^i \nabla_j(fc_3\zeta^j)
\right.
\\ &
\left.
+ \frac{2\zeta^i}{\hat T^2} \epsilon^{klm} \zeta_k \left[\frac{\del g_1}{ \del \psi} \del_l A_m
+  T_0\frac{\del g_2}{ \del \psi} \del_l a_m \right]
+g_1 \epsilon^{ijk} \del_j A_k + g_2 T_0 \epsilon^{ijk} \del_j a_k
\right)\ ,
\end{split}
\end{equation}
where we have used $H_a=(\hat T,\hat \nu,\zeta^2)$ for $a=1\dots 3$. The derivative is covariant w.r.t to the three dimensional metric.

In the next appendix we will try and solve this equation. It should be noted that in general for non-local terms the derivative expansion fails. But since all our Kubo formulas will be evaluated with momenta directed along one of the axes only, in our case we can still rely on the consistency of an expansion in powers of momenta (momenta in numerator and denominator must either cancel or vanish).

\section{Solving the Goldstone E.O.M for Non-Local Terms}\label{app:solving_eom}
In this appendix we want to solve the Goldstone equation of motion for the expectation value of the Goldstone phase $\phi$ defined through:
\begin{equation}\label{app:eq:solving_eom:zeta_eq_form_phi}
\zeta_{eq}^i=A^i-\del^i\phi \ .
\end{equation}
We will do this in two steps. First we will solve the E.O.M at lowest order in derivatives:
\begin{equation}\label{app:eq:solving_eom:zeroth_eom}
\nabla_i \left(\frac{f}{T}\zeta_{eq}^i\right)=0.
\end{equation}
Then we will add the next order derivative corrections \eqref{app:eq:Goldsteom:eom_supf_full} to the E.O.M and correct our solution accordingly.

We can solve the E.O.M order by order in the variation of the sources.
For our Kubo formulas we only need to solve up to first order in the metric and gauge field perturbation.
This is due to the fact that all our Kubo formulas are given in terms of two point function.
We will not be interested in correlators including spatial components of the stress tensor.
We may therefore immediately set $g_{ij}=\delta_{ij}$.
Let us expand \eqref{app:eq:solving_eom:zeroth_eom} to linear order in the other external sources:\footnote{Remember that in the absence of sources $\vec\zeta_{eq}=\vec \zeta_0$ is constant so $\phi$ is of first or higher order in the variation of external sources.}
\begin{equation}
\begin{split}
\nabla_i & \left(\frac{f}{T}\zeta_{eq}^i \right)  =
\del_i \left(\frac{f\left(T_0 e^{-\sigma},\frac{\mu_0+\delta A_0}{T_0},\zeta_{eq}^2\right)}{T_0e^{-\sigma}} \zeta_{eq}^i\right) 
\\ & =
    \frac{1}{T_0}\del_i
    \left(f_0 \zeta_{0}^i\right)
    +\frac{\zeta_0^i}{T_0} \del_i  \left(f_0\sigma
    -T_0 (f_T)_0 \sigma +(f_\nu)_0 \frac{\delta A_0}{T_0}
    +2 (f_{\zeta^2})_0 \zeta_0^k \left(\delta A_k-\del_k \phi\right) \right)
    \\
    &\ \ \ +\frac{f_0}{T_0}\del_i
    \left(\delta A^i-\del^i \phi\right)
+ O(\delta^2)\\ &
    =
    \frac{f_0}{T_0}\del_i
    \left(\delta A^i-\del^i \phi\right)
    +\frac{f_0-T_0(f_T)_0}{T_0}\zeta_0^i \del_i  \sigma
    +\frac{(f_\nu)_0}{T_0^2} \zeta_0^i \del_i \delta A_0
    \\
    &
    \ \ \ +\frac{2(f_{\zeta^2})_0}{T_0}\zeta_0^i \zeta_0^k \del_i \left(\delta A_k-\del_k \phi\right)
    + O(\delta^2)
\\ &
    =
    -\frac{f_0}{T_0}\del^2 \phi-\frac{2(f_{\zeta^2})_0}{T_0} (\zeta_0 \cdot \del)^2 \phi+
    \frac{f_0}{T_0}\del_i \delta A^i +\frac{2(f_{\zeta^2})_0}{T_0}\zeta_0^k (\zeta_0 \cdot  \del )\delta A_k
    \\
    &
    \ \ \
    +\frac{f_0-T_0(f_T)_0}{T_0}(\zeta_0 \cdot \del)  \sigma
    +\frac{(f_\nu)_0}{T_0^2} (\zeta_0 \cdot \del) \delta A_0
    + O(\delta^2)
=0\ ,
\end{split}
\end{equation}
where we have used $\zeta_{eq}^i = \zeta_0^i+ \delta A^i-\del^i \phi$, $\delta A_0= A_0-\mu_0$ and $f_0=f\left(T_0,\frac{\mu_0}{T_0},\zeta_0^2\right)$ and similarly for the derivatives $f_T,~f_\nu,~f_{\zeta^2}$ w.r.t $T,~\nu$ and $\zeta^2$.
Integrating out the Goldstone mode amounts to solving this equation for the expectation value of the Goldstone phase $\langle\phi\rangle$ and plugging back the solution into the current or stress tensor.

Solving in momentum space we get:
\begin{equation}
\begin{split}&
 \langle \phi \rangle= -\frac{
    i f_0 k_i \delta A^i + i (\zeta_0 \cdot k ) \left[2(f_{\zeta^2})_0  (\zeta_0 \cdot \delta A)
    +(f_0-T_0(f_T)_0) \sigma
    +\frac{(f_\nu)_0}{T_0}  \delta A_0 + \frac{f_0}{2}  \delta g_{ii} \right]}
    {f_0 k^2+2(f_{\zeta^2})_0 (\zeta_0 \cdot k)^2}
\end{split}
\end{equation}
up to linear order of the sources.\footnote{This is at order $-1$ in momentum. Any higher momentum correction to the equation of motion can add a correction to $\phi$ of order $0$ in momentum or higher. $\vec\zeta_{eq}$ will be corrected at order $1$ or higher in momentum.}
we therefore have in momentum space:
\begin{equation}\label{app:eq:zeta_momentum}
\zeta^i_{eq} = \zeta_0^i+\delta A^i-ik^i\langle\phi\rangle \ .
\end{equation}

Repeating the same analysis for the corrected E.O.M \eqref{app:eq:Goldsteom:eom_supf_full} we get:
\begin{equation}
\begin{split}\label{app:eq:full_phi_sol}
  \langle \phi \rangle = & -\frac{1} {f k^2+2 f_{\zeta^2} (\zeta_0 \cdot k)^2} \times \\ &
    \qquad  \biggr[
    2 (\zeta \cdot \delta A) \left[fT_0c_3k^2 +i f_{\zeta^2} \left(\zeta_0 \cdot k\right)\right]
    +if (k\cdot \delta A) (1+2 ic_3T_0 (\zeta_0\cdot k))
    \\& \qquad \qquad
    +\frac{2}{T_0} \frac{\del g_1}{\del \psi} (\zeta_0\cdot k) \epsilon^{klm}\zeta^0_k k_l \delta A_m
    +2 \frac{\del g_2}{\del \psi} (\zeta_0\cdot k) \epsilon^{klm}\zeta^0_k k_l  a_m
    \\ & \qquad
    +\sigma \left[
    -fc_1T_0^2 k^2
    -2T_0^2(\zeta_0\cdot k)^2 \left(\frac{\del(fc_1)}{\del\zeta^2}-\frac{f}{T}\frac{\del (T c_3)}{\del T}\right)
    \right.
    \\  &   \qquad \qquad
    -i(T_0f_T-f) (\zeta_0 \cdot k) (1+2 i c_3T_0(\zeta_0\cdot k))
     \biggr]
    \\ &  \qquad
    +\delta A_0 \left[f c_2 k^2 +i\frac{f_\nu}{T_0} (\zeta_0\cdot k) \left(1+2iT_0c_3\left(\zeta_0\cdot k\right)\right)
    \right.
    \\
    & \qquad \qquad
    \left.\left.+ 2 (\zeta_0\cdot k)^2 \left(-f\frac{\del{c_3}}{\del\nu}+\frac{\del(fc_2)}{\del\zeta^2}\right)
    \right]\right] + O(\delta^2,k)
\end{split}
\end{equation}
all the $c_i, g_i, f$ and their derivatives are evaluated in terms of flat space parameters $(T_0,\nu_0,\zeta_0^2)$.

This is the expectation value for the field $\phi$. Differentiating w.r.t the various sources and setting the sources to zero we will be able express $c_1$,$c_2$ and $c_3$ in terms of correlation functions of the Goldstone phase gradient and another (composite) operator. In the special case $\vec \zeta_0 \perp \vec k$ we have a simpler expression:
\begin{equation}
\begin{split}\label{app:eq:perp_phi_sol}
 \langle \phi \rangle = & - i\frac{k\cdot \delta A}{k^2}
 - 2 (\zeta \cdot\delta  A) T_0c_3
     + \sigma c_1T_0^2
     -\delta A_0  c_2\ ,
\end{split}
\end{equation}
whereas for $\vec \zeta_0 \parallel \vec k$ we have:
\begin{equation}
\begin{split}\label{app:eq:paral_phi_sol}
 \langle \phi \rangle = & -i\frac{k\cdot\delta A}{k^2}\ .
\end{split}
\end{equation}

\section{First Order Charged Fluid Dynamics in 2+1 Dimensions}\label{sec:2+1}
In this appendix we use our method to rederive Kubo formulas for a 2+1 dimensional parity violating charged fluid up to first order in the derivative expansion.

\subsection{Preliminaries}
The most general partition function for such a fluid is given in terms of two thermodynamical functions $\alpha$ and $\beta$ as follows:
\begin{align}
\begin{split}\notag
\ln Z = W^0 +W^1
\end{split}\\
\begin{split}
W^0 = \int d^2x \sqrt{g_2} \frac{e^\sigma}{T_0} P(T_0 e^{-\sigma}, e^{-\sigma}A_0) ,
\end{split}\\
\begin{split}\notag
W^1 = \frac{1}{2} \int \left(\alpha(\sigma,A_0)dA + T_0 \beta(\sigma,A_0) da\right) ,
\end{split}
\end{align}
where
\begin{equation}
\frac{1}{2} \int dY = \int d^2 x \sqrt{g_2} \epsilon^{ij} \del_i Y_j ,
\end{equation}
and the dependence of $\alpha$ and $\beta$ on $T_0$ is hidden in their $\sigma,A_0$ dependence as follows \cite{Banerjee:2012iz}:
\begin{equation} \label{2+1:Tmu_dependencies}
W(\sigma,A_0,a_i,A_i,g^{ij}) = \mathcal{W} (\frac{e^\sigma}{T_0},\frac{A_0}{T_0},T_0a_i,A_i,g^{ij})\ .
\end{equation}

Using equations \eqref{anom:JT_formulas} one is able to extract expressions for the stress tensor and charge current (up to first order in the derivative expansion) consistent with this partition function\cite{Banerjee:2012iz}:
\begin{align}
\begin{split}\label{2+1:Tij}
T^{ij}=Pg^{ij},
\end{split}\\
\begin{split}
T_{00} = -e^{2\sigma}\left(P-aP_a-bP_b\right)
         - T_0e^\sigma \left(\frac{\del \alpha}{\del \sigma} \epsilon^{ij} \del_i A_j
         +  T_0 \frac{\del\beta}{\del\sigma} \epsilon^{ij}\del_i a_j\right),
\end{split}\\
\begin{split}\label{2+1:T0i}
T_0^i = T_0 e^{-\sigma} \left(
\left(T_0 \frac{\del \beta}{\del\sigma} - A_0 \frac{\del\alpha}{\del\sigma}\right)\epsilon^{ij}\del_j\sigma
+ \left(T_0 \frac{\del \beta}{\del A_0} - A_0 \frac{\del\alpha}{\del A_0}\right)\epsilon^{ij}\del_j A_0
\right),
\end{split}\\
\begin{split}
J_0 = -e^{\sigma} P_b - T_0e^\sigma \left(\frac{\del \alpha}{\del A_0} \epsilon^{ij} \del_i A_j
         +  T_0 \frac{\del\beta}{\del A_0} \epsilon^{ij}\del_i a_j\right),
\end{split}\\
\begin{split}\label{2+1:Ji}
J^i = T_0e^{-\sigma} \left(\frac{\del \alpha}{\del \sigma} \epsilon^{ij} \del_j \sigma
         +  \frac{\del\alpha}{\del A_0} \epsilon^{ij}\del_j A_0 \right).
\end{split}
\end{align}

\subsection{Extracting the Kubo Relations}
We can now differentiate equations \eqref{2+1:Tij}-\eqref{2+1:Ji} with respect to the various different sources (using the independent set of variables from equation \eqref{anom:indep_var}, and the differentiation rules from \eqref{anom:JT_formulas}). After setting the metric and gauge field perturbation to zero we get the following Kubo relations (in momentum space):
\begin{align}
\begin{split}\label{2+1:Kubo2+1:1}
i\lim_{k\rightarrow 0}\frac{\epsilon_{ij} k^j}{k^2} G^{0i,00}(k,-k)\biggr|_{\omega=0} = - T_0
    \left(T_0\frac{\del\beta}{\del\sigma}- \mu_0\frac{\del\alpha}{\del\sigma}\right)\biggr|_{\sigma=\mathcal{A}_0=0},
\end{split}\\
\begin{split}
i\lim_{k\rightarrow 0}\frac{\epsilon_{ij} k^j}{k^2} G^{i,00}(k,-k)\biggr|_{\omega=0} = T_0  \frac{\del\alpha}{\del\sigma}\biggr|_{\sigma=\mathcal{A}_0=0},
\end{split}\\
\begin{split}
i\lim_{k\rightarrow 0}\frac{\epsilon_{ij} k^j}{k^2} G^{0i,0}(k,-k)\biggr|_{\omega=0} = T_0
    \left(T_0\frac{\del\beta}{\del A_0}- \mu_0\frac{\del\alpha}{\del A_0}\right)\biggr|_{\sigma=\mathcal{A}_0=0},
\end{split}\\
\begin{split}\label{2+1:Kubo2+1:4}
i\lim_{k\rightarrow 0} \frac{\epsilon_{ij} k^j}{k^2} G^{i,0}(k,-k)\biggr|_{\omega=0} = -T_0 \frac{\del\alpha}{\del A_0}\biggr|_{\sigma= \mathcal{A}_0=0},
\end{split}
\end{align}
where we used $J^0 = (J_0-g_{0i}J^i)/g_{00}$ and $T^{00}=(T_{00}-2g_{0i}T_0^i+g_{0i}g_{0j}T^{ij})/g_{00}^2$ to relate the upper temporal components of the stress tensor and current to the known components given by equations \eqref{2+1:Tij}-\eqref{2+1:Ji}. After differentiating and setting the background perturbation to zero only the first term of each of these expressions survives.

Bearing in mind the implicit $T_0$ dependence of $\alpha$ and $\beta$ (see \eqref{2+1:Tmu_dependencies}):
\begin{equation}\label{2+1:alpha_dep}
\alpha(\sigma,A_0) \equiv \widetilde\alpha (T_0 e^{-\sigma},\frac{A_0}{T_0}) = \widetilde\alpha (T,\nu \small\equiv \frac{\mu}{T} )
\end{equation}
where $T,\mu$ are the local values of the temperature and chemical potential at zero (and as turns out from \cite{Banerjee:2012iz} also first) order in the derivative expansion, we can recast \eqref{2+1:Kubo2+1:1}-\eqref{2+1:Kubo2+1:4} as:
\begin{align}
\begin{split}\label{2+1:ab1}
  \boxed{ \frac{\del\tilde\alpha}{\del \nu}\biggr{)}_T = -i\lim_{k\rightarrow 0} \frac{\epsilon_{ij} k^j}{k^2} G^{i,0}(k,-k)\biggr|_{\omega=0}}\ ,
\end{split}\\
\begin{split}
\boxed{\frac{\del\tilde \alpha}{\del T}\biggr{)}_\nu = -\frac{i}{T^2}\lim_{k\rightarrow 0}\frac{\epsilon_{ij} k^j}{k^2} G^{i,00}(k,-k)\biggr|_{\omega=0}}\ ,
\end{split}\\
\begin{split}
\boxed{\frac{\del\tilde\beta}{\del \nu} \biggr{)}_T = \nu\frac{\del\tilde\alpha}{\del \nu}+
 \frac{i}{T}\lim_{k\rightarrow 0}\frac{\epsilon_{ij} k^j}{k^2} G^{0i,0}(k,-k)\biggr|_{\omega=0}} \ ,
\end{split}\\
\begin{split}\label{2+1:ab4}
\boxed{ \frac{\del\tilde\beta}{\del T}\biggr{)}_\nu = \nu \frac{\del\tilde\alpha}{\del T}+
  \frac{i}{T^3}\lim_{k\rightarrow 0}\frac{\epsilon_{ij} k^j}{k^2} G^{0i,00}(k,-k)\biggr|_{\omega=0}}\ .
\end{split}
\end{align}
These fully determine $\alpha$ and $\beta$ (up to a temperature/chemical potential independent constant).
The correlators on the RHS are calculated in flat space where $T=T_0$, $A_0=\mu=\mu_0$. In curved (stationary) space the only difference in the transport functions $\alpha$ and $\beta$ would be changing $T_0,\mu_0\rightarrow T,\mu$. No further metric and gauge field dependence can be introduced into the thermal transport functions because of \eqref{2+1:alpha_dep}.

\subsection{Hydrodynamic Transport Coefficients}
After using the equilibrium partition function to derive expressions for the stress-tensor and charge current in equilibrium (Eqs.~\eqref{2+1:Tij}-\eqref{2+1:Ji}), the authors of \cite{Banerjee:2012iz} compared them to their most general hydrodynamic allowed form:
\begin{align}
\begin{split}
T^{\mu\nu} = \epsilon u^\mu u^\nu + P P^{\mu\nu} - \eta \sigma^{\mu\nu} - \tilde\eta \tilde\sigma^{\mu\nu} - P^{\mu\nu} \zeta \nabla_\alpha u^\alpha - P^{\mu\nu} \left( \tilde\chi_B B+ \tilde\chi_\Omega \Omega \right),
\end{split}\\
\begin{split}
J^\mu = \rho u^\mu +\sigma V^\mu +\tilde\sigma \tilde V^\mu + \tilde\chi_E \tilde E^\mu +\tilde \chi_T \tilde T^\mu,
\end{split}
\end{align}
evaluated on the most general equilibrium solution.
The conventions are as follows: $\sigma^{\mu\nu} = P^{\mu\alpha} P^{\nu\beta}(\nabla_\alpha u_\beta +\nabla_\beta u_\alpha - g_{\alpha\beta} \nabla_\lambda u^\lambda)$,
$\tilde\sigma^{\mu\nu} = \epsilon^{\alpha\rho(\mu}u_\alpha \sigma_\rho^{\nu)}$,
$B = -\frac{1}{2}\epsilon^{\mu\nu\rho} u_\mu \mathcal{F}_{\nu\rho}$,
$\Omega = -\epsilon^{\mu\nu\rho} u_\mu \nabla_\nu u_\rho$,
$V^\mu = E^\mu - T P^{\mu\nu}\nabla_\nu\frac{\mu}{T}$,
$\tilde V^\mu = \epsilon ^{\mu\nu\rho} u_\nu V_\rho$,
$E^\mu = \mathcal F^{\mu\nu} u_\nu$, $\tilde E^\mu = \epsilon^{\mu\nu\rho} u_\nu E_\rho$,
and  $\tilde T^\mu = \epsilon^{\mu\nu\rho} u_\nu \nabla_\rho T$.

This comparison allowed the authors of \cite{Banerjee:2012iz} to express the thermal transport coefficients $\tilde\chi_B, \tilde\chi_\Omega, \tilde\chi_E, \tilde\chi_T$ (those that affect the fluid's behavior in equilibrium) in terms of the thermal functions $\alpha,\beta$:
\begin{align}
\begin{split}
\tilde\chi_B &= \frac{\del P}{\del \epsilon} \left( - T \frac{\del \alpha}{\del \sigma } \right)
+ \frac{\del P} {\del \rho} \left(T_0 \frac{\del \alpha}{\del A_0}\right)\ ,
\end{split}\\
\begin{split}
\tilde\chi_\Omega &= \frac{\del P}{\del \epsilon}  T \left( T\frac{\del\beta}{\del\sigma} - \mu \frac{\del\alpha}{\del\sigma}\right)
- \frac{\del P} {\del \rho} T \left(T_0 \frac{\del \beta}{\del A_0} -A_0 \frac{\del \alpha}{\del A_0}\right)\ ,
\end{split}\\
\begin{split}
\tilde\chi_E &= T_0 \frac{\del\alpha}{\del A_0} + \frac{\rho}{\epsilon+P} T \left( T_0\frac{\del\beta}{\del A_0} - A_0 \frac{\del\alpha}{\del A_0}\right)\ ,
\end{split}\\
\begin{split}
T \tilde\chi_T &=
-T \frac{\del\alpha}{\del \sigma} - \frac{\rho}{\epsilon+P} T \left( T\frac{\del\beta}{\del \sigma} - \mu \frac{\del\alpha}{\del \sigma}\right)\ .
\end{split}
\end{align}

Using the Kubo formulas we found for $\alpha$ and $\beta$ in Eqs.~\eqref{2+1:ab1}-\eqref{2+1:ab4} we can immediately present these in form of Kubo-formulas for the non-dissipative transport coefficients $\tilde\chi_B, \tilde\chi_\Omega, \tilde\chi_E, \tilde\chi_T$:
\begin{align}
\begin{split}
\boxed{\tilde\chi_B = i\lim_{k\rightarrow 0}\frac{\epsilon_{ij} k^i}{k^2} \left[ \frac{\del P}{\del \epsilon} G^{j,00}(k,-k)
+\frac{\del P} {\del \rho} G^{j,0}(k,-k) \right]\biggr|_{\omega=0} }\ ,
\end{split}\\
\begin{split}
\boxed{\tilde\chi_\Omega =i\lim_{k\rightarrow 0}\frac{\epsilon_{ij} k^i}{k^2} \left[ \frac{\del P}{\del \epsilon} G^{0j,00}(k,-k) + \frac{\del P} {\del \rho}  G^{0j,0}(k,-k) \right]\biggr|_{\omega=0}}\ ,
\end{split}\\
\begin{split}
\boxed{\tilde\chi_E =i\lim_{k\rightarrow 0} \frac{\epsilon_{ij} k^i}{k^2} \left[ G^{j,0}(k,-k)
- \frac{\rho}{\epsilon+P} G^{0j,0}(k,-k) \right]\biggr|_{\omega=0}}\ ,
\end{split}\\
\begin{split}
\boxed{T \tilde\chi_T=i\lim_{k\rightarrow 0}\frac{\epsilon_{ij} k^i}{k^2} \left[ G^{j,00}(k,-k)
- \frac{\rho}{\epsilon+P} G^{0j,00}(k,-k) \right] \biggr|_{\omega=0} }\ ,
\end{split}
\end{align}
which reproduces the results of \cite{Jensen:2011xb} (Eq.~(1.10)) using this simpler derivation. The minus sign differences are due to a different definition of the Green functions. 
Our Green function indexes are inverted compared to those of \cite{Jensen:2011xb} see their  Eq.~(4.2). 
This can be compensated by taking $k\rightarrow-k$ in some of the Kubo formulas, thus inducing a minus sign.

\section{3 + 1 Dimensional Uncharged Fluid at Second Order in Derivatives}\label{sec:second_order}
In this appendix we use our method to rederive Kubo formulas for a 3 + 1 dimensional neutral fluid (including parity violating contributions) up to second order in derivatives. Parity violating terms in the stress-tensor and charge-current were considered (and dismissed) by the calculation of\cite{Banerjee:2012iz}. This is due to the fact that the only parity odd possible contribution to the partition function turns out to be a total derivative term.

\subsection{Preliminaries}
The most general equilibrium partition function for the fluid described above is given by:
\begin{gather}\label{3+1-2nd:partition}
\begin{split}
& \ln Z = W^0+W^2 \\
& W^0 = \int \sqrt{g_3} \frac{e^\sigma}{T_0} P(T_0 e^{-\sigma}), \\
& W^2 = - \frac{1}{2} \int d^3 x \sqrt{g_3} \left[\tilde{P}_1(T_0 e^{-\sigma}) R +T_0^2 \tilde{P}_2(T_0 e^{-\sigma}) f_{ij} f^{ij} + \tilde{P}_3 (T_0 e^{-\sigma}) (\del \sigma)^2 \right],
\end{split}
\end{gather}
where the (zeroth order) local value of the temperature is $T\equiv T_0 e^{-\sigma}$ (formerly denoted $a$), $R$ is the Ricci scalar of the 3 dimensional metric $g_{ij}$, $f_{ij}=\del_i a_j - \del_j a_i$ and we shall often use $P_i (\sigma) \equiv \tilde{P}_i (T_0 e^{-\sigma})$.

Using the uncharged analog of eq.\eqref{anom:JT_formulas} the authors of \cite{Banerjee:2012iz} were able to find the stress-tensor components:
\begin{align}
\begin{split}\label{3+1-2nd:T_ij}
T^{ij}&=Pg^{ij}
+TP_1\left(R^{ij}-\frac{1}{2}Rg^{ij}\right) + 2 T_0^2 T P_2 \left(f^{ik}f^j{}_k -\frac{1}{4}f^2g^{ij}\right)
+\frac{1}{2} T P''_1 (\nabla \sigma)^2 g^{ij}
\\
&+T(P_3-P_1'')\left(\nabla^i\sigma \nabla^j \sigma - \frac{1}{2} (\nabla \sigma)^2 g^{ij}\right)
-TP'_1 (\nabla^i\nabla^j \sigma -g^{ij}\nabla^2\sigma )
\end{split}\\
\begin{split}
T_{00} &= - e^{2\sigma}\left(P-TP_T\right)
+\frac{T_0^2}{2T} \left(P'_1R+T_0^2 P'_2f^2 - P'_3 (\nabla \sigma)^2 -2 P_3\nabla^2 \sigma \right),
\end{split}\\
\begin{split}\label{3+1-2nd:T0i}
T_0^i & =  2T_0^2 T (P'_2 \nabla_j \sigma f^{ji} + P_2 \nabla_j f^{ji}),
\end{split}
\end{align}
where ' denotes derivatives with respect to $\sigma$, $T$ subscript denotes derivatives with respect to the zeroth order temperature $T=T_0 e^{-\sigma}$, $\nabla$ is the covariant 3-derivative and $R$ stands for the \emph{three dimensional} Ricci Tensor/Scalar of $g_{ij}$.

\subsection{Extracting the Kubo Relations}
To extract the Kubo relations, one has to vary equations \eqref{3+1-2nd:T_ij}-\eqref{3+1-2nd:T0i} with respect to the various sources. Some of the Kubo relations we present in this section include three point functions.
Because of this reason, using the set of independent variables of equations \eqref{anom:indep_var}-\eqref{anom:JT_formulas} will involve multiple instances of raising/lowering indexes, as well as careful surveillance of the point at which the differentiation is carried out. This encouraged us to use $\delta g_{\mu\nu}=g_{\mu\nu}- \eta_{\mu\nu} \equiv h_{\mu\nu}$ as the independent set of variables instead, differentiating according to \eqref{anom:TJ_orig} directly. Differentiating according to \eqref{anom:indep_var}-\eqref{anom:JT_formulas} accompanied by a careful bookkeeping of indexes and momenta gives precisely the same results.

To differentiate with respect to $h_{\mu\nu}$ we have to express the stress tensor components \eqref{3+1-2nd:T_ij}-\eqref{3+1-2nd:T0i} as functions of $h_{\mu\nu}=g_{\mu\nu}- \eta_{\mu\nu}$ rather than $\sigma, a_i, g^{ij}$. For instance, to replace $\sigma$ with $h_{tt}$ one can use $\sigma = \frac{1}{2} \ln (1-h_{tt}) = - \frac{1}{2} (h_{tt}+h_{tt}^2/2) + O (h^3)$.
Here and in what follows we replace sub/superscripts $(0,1,2,3)$ with $(t,x,y,z)$.
Similar relations allow us to express $a_i$ and $g^{ij}$ as a function of the various components of $h_{\mu\nu}$. Plugging these expressions into \eqref{3+1-2nd:T_ij}-\eqref{3+1-2nd:T0i} gets us to our starting point of our Kubo formula analysis. We have revealed the full dependence of the stress tensor on the metric perturbation without having to solve the equation of motion for the fluid velocity and temperature first. Our analysis follows closely the one in \cite{Moore:2012tc}, significantly shortened by using the results of \cite{Banerjee:2012iz}.

Without further ado let us start with the Kubo formula for $\tilde P_1 (T)$. For this purpose we vary $T^{xy}$ with respect to $h_{xy}(z)$. This metric perturbation is related to our metric variables as follows $\sigma=0$, $a_i = 0$, $g_{xy}=h_{xy}$. Plugging this perturbation into \eqref{3+1-2nd:T_ij} we obtain:
\begin{equation}
T^{xy} = -Ph_{xy} -\frac{1}{2} TP_1 \frac{\del^2 h_{xy}}{\del z^2} + O(h^2).
\end{equation}
Performing the variation, then setting (what's left of) the metric perturbation to zero we get in momentum space:
\begin{equation}\label{3+1-2nd:P_1}
\boxed{\tilde P_1 (T) = \frac{1}{T} \lim_{k_z \rightarrow 0 } \frac{\del^2}{\del k_z^2} G^{xy,xy} (k,-k) \biggr|_{k_0=0}}
\ .
\end{equation}
The correlation function is calculated in flat space with temperature $T_0$, the full $T=T_0e^{-\sigma}$ dependence is easily restored by replacing $T_0$ with $T$.

To obtain a Kubo formula for $\tilde P_2 (T)$ we vary $T^{xy}$ with respect to both $h_{xt}(z)$ and $h_{yt}(z)$. The corresponding $\sigma,a_i,g_{ij}$ are $\sigma=0$, $a_i = -h_{it}$, $g_{ij}=\delta_{ij} + h_{it}h_{jt}$. plugging this perturbation into \eqref{3+1-2nd:T_ij} we get:
\begin{equation}
T^{xy} = -Ph_{xt}h_{yt} + (-TP_1+2T^3P_2) \del_z h_{xt} \del_z h_{yt} - \frac{TP_1}{2} (h_{xt} \del^2_z h_{yt}+h_{yt} \del^2_z h_{xt}) +O(h^3)\ .
\end{equation}
Performing the variation, then setting the metric perturbation to zero we get in momentum space:
\begin{equation}\label{3+1-2nd:P_2}
\boxed{\tilde P_2 (T)=\frac{\tilde P_1}{2T^2} - \frac{1}{2T^3}\lim_{p_z,q_z \rightarrow 0 } \frac{\del^2}{\del p_z \del q_z} G^{xt,yt,xy} (p,q,-p-q) \biggr|_{p_0=q_0=0} }\ .
\end{equation}

Last but not least, we derive a Kubo formula for $\tilde P_3 (T)$ by varying $T^{xy}$ with respect to $h_{tt}(x,y)$ (twice). Our metric variables then become $\sigma=- \frac{1}{2} (h_{tt}+h_{tt}^2/2) + O (h^2)$, $a_i = 0$, $g_{ij}=\delta_{ij}$. plugging this perturbation into \eqref{3+1-2nd:T_ij} we get:
\begin{equation}
T^{xy} = T^{xy}_h + \frac{T}{4} (P_3-P''_1+2P'_1) \del_x h_{tt} \del_y h_{tt} +\frac{T}{2} P'_1 h_{tt} \del_x \del_y h_{tt} +O(h^3)\ .
\end{equation}
Performing the variation, then setting the metric perturbation to zero we get in momentum space:\\
\begin{equation}\label{3+1-2nd:P_3}
  \addtolength{\fboxsep}{5pt}
   \boxed{
   \begin{gathered}
      \tilde P_3 (T)= P''_1-2P'_1 - \frac{1}{T} \lim_{p_x,q_y \rightarrow 0 } \frac{\del^2}{\del p_x \del q_y} G^{tt,tt,xy} (p,q,-p-q) \biggr|_{p_0=q_0=0}\\
      =T^2P_{1 TT} + 3TP_{1T} - \frac{1}{T} \lim_{p_x,q_y \rightarrow 0 } \frac{\del^2}{\del p_x \del q_y} G^{tt,tt,xy} (p,q)
   \end{gathered}
   }\ .
\end{equation}
Note that our definition of the Green function implies a factor of 2 when varying with respect to a perturbation of two identical indexes. No extra factor of 2 originates from the differentiation of $\del_x h_{tt}\del_y h_{tt}$ w.r.t $h_{tt}$. This is due to the fact that differentiating this term twice w.r.t $h_{tt}$ we get in momentum space a contribution proportional to $p_x q_y +p_y q_x$. Only the first of these contributes to  $\frac{\del^2}{\del p_x \del q_y}$\ .

\subsection{Hydrodynamic Transport Coefficients}
The hydrodynamic transport coefficients for a 3+1 dimensional neutral fluid at 2nd order in the derivative expansion are defined by:
\begin{align}
\begin{split}
T^{\mu\nu}_{(2)}=\
&T
\left[
\kappa_1\tilde R_{\langle\mu\nu\rangle}
-\kappa_2 \tilde R_{\alpha \langle\mu\nu\rangle \beta} u^\alpha u^\beta
+\lambda_3\omega^\alpha{}_{\langle\mu} \omega_{\nu\rangle\alpha}
+\lambda_4 a_{\langle\mu}a_{\nu\rangle}\right]
\\
+ \ & T P_{\mu\nu}  \left[\zeta_2 \tilde R + \zeta_3 \tilde R_{\mu\nu}u^\mu u^\nu
-\xi_3 \omega_{\mu\nu} \omega^{\mu\nu} +\xi_4 a^2
\right]
+{\mbox{\small dissipative}\atop \mbox{\small contributions}} \ ,
\end{split}
\end{align}
where $\tilde R$ is the 4-dimensional Riemann tensor, $\omega^{\mu\nu} = P^{\mu\alpha} P^{\nu\beta} \left(\frac{\nabla_\alpha u_\beta - \nabla_\beta u_\alpha}{2}\right)$ is the vorticity tensor, $a^\mu=u\cdot\nabla u^\mu$ is the acceleration vector and all the coefficients are as yet arbitrary functions of the temperature. Angular brackets denote the traceless symmetrized transverse projection of any tensor.

Using the expressions found in \cite{Banerjee:2012iz} (equations (5.8) and (5.15)) for the non-dissipative hydrodynamic transport coefficients ($\kappa_1,\kappa_2,\lambda_3,\lambda_4,\zeta_2,\zeta_3,\xi_3,\xi_4$) in terms of the thermal functions $P_1,P_2,P_3$, and plugging the thermal function in terms of their Kubo formulas as found in the previous subsection we find the following Kubo formulas for the hydrodynamic transport coefficients:
\begin{equation}
  \addtolength{\fboxsep}{5pt}
   \boxed{
   \begin{gathered}
        \begin{split}
            \kappa_1 = \ & P_1 = \frac{1}{T} \lim_{k_z \rightarrow 0 } \frac{\del^2}{\del k_z^2} G^{xy,xy} (k,-k)
            \biggr|_{k_0=0} , \\
            \lambda_3 =\  & -8T^2P_2 - P'_1 +3P_1 \\
            = \ & - \kappa_1 + T \frac{d\kappa_1}{dT} + \frac{4}{T} \lim_{p_z,q_z \rightarrow 0 } \frac{\del^2}{\del p_z \del q_z}
            G^{xt,yt,xy} (p,q) \biggr|_{p_0=q_0=0},\\
            \lambda_4 = \ & P_3-P''_1+P'_1 \\
            = \ &  T \frac{d\kappa_1}{dT} -\frac{1}{T}  \lim_{p_x,q_y \rightarrow 0 } \frac{\del^2}{\del p_x \del q_y}
            G^{tt,tt,xy} (p,q)\biggr|_{p_0=q_0=0}\ .
        \end{split}
  \end{gathered}
 }
\end{equation}
All the other transport terms are given in terms of $\kappa_1$, $\lambda_3$ and $\lambda_4$ by the following relations:
\begin{equation}
  \addtolength{\fboxsep}{5pt}
   \boxed{
   \begin{gathered}
        \begin{split}
            \kappa_2 =\ & \kappa_1+T\frac{d\kappa_1}{dT}\ , \\
            \zeta_2 =\ & \frac{1}{2}\left[s\frac{d\kappa_1}{ds}-\frac{\kappa_1}{3}\right]\ ,\\
            \zeta_3 = \ & \left(s\frac{d\kappa_1}{ds}+\frac{\kappa_1}{3} \right)
            +\left(s\frac{d\kappa_2}{ds}-\frac{2\kappa_2}{3} \right) +
            \frac{s}{T}\left(\frac{dT}{ds}\right)\lambda_4\ ,\\
            \xi_3 =\ &  \frac{3}{4}\left(\frac{s}{T}\right)\left(\frac{dT}{ds}\right)
            \left(T\frac{d\kappa_2}{dT}+2\kappa_2 \right) - \frac{3\kappa_2}{4}+
            \left(\frac{s}{T}\right)\left(\frac{dT}{ds}\right)\lambda_4\
            \\ & +\frac{1}{4} \left[s\frac{d\lambda_3}{ds}+\frac{\lambda_3}{3}
            -2 \left(\frac{s}{T}\right)\left(\frac{dT}{ds}\right)\lambda_3 \right]\ ,
            \\
            \xi_4 =\ & -\frac{\lambda_4}{6}
            -\frac{s}{T} \left(\frac{dT}{ds}\right)
                \left(\lambda_4+\frac{T}{2}\frac{d\lambda_4}{dT} \right)
            - T\left(\frac{d\kappa_2}{dT}\right)\left(\frac{3s}{2T}\frac{dT}{ds}-\frac{1}{2}\right)\\
            &\ -\frac{Ts}{2}\left(\frac{dT}{ds}\right)\left(\frac{d^2\kappa_2}{dT^2}\right)
            \ .
        \end{split}
  \end{gathered}
 }\
\end{equation}
After translating between the different conventions for the transport coefficients, these match the Kubo formulas of \cite{Moore:2012tc}  Eqs.~(2.10)-(2.17) found here in this shorter derivation. Small differences in our expression for $\lambda_4$ compared to  Eq.~(2.12) of \cite{Moore:2012tc}, are due to slight typos in their derivation in the appendix (A.4). 
\subsection{New Identities among correlation functions}
The Kubo formula we derived for the thermal functions Eqs.~(\ref{3+1-2nd:P_1}, \ref{3+1-2nd:P_2}, \ref{3+1-2nd:P_3}) are not unique.
For instance $P_2$ can be obtained both from the three point function as we derived in \eqref{3+1-2nd:P_2}, as well as from the two point function of $T^{tx} T^{ty}$:

\begin{equation}
P_2 (T) = \frac{1}{2T^3}  \lim_{k_x,k_y \rightarrow 0 } \frac{\del^2}{\del k_x \del k_y} G^{tx,ty} (k,-k)\biggr|_{k_0=0}\ .
\end{equation}
This implies the following identity:
\begin{equation}
 \boxed{ \lim_{\vec{k} \rightarrow 0 }
\left[\frac{\del^2}{\del k_z^2} G^{xy,xy} (k) - \frac{\del^2}{\del k_x \del k_y} G^{tx,ty} (k) \right]
=\lim_{p_z,q_z \rightarrow 0 } \frac{\del^2}{\del p_z \del q_z} G^{xt,yt,xy} (p,q)}\ .
\end{equation}
We do not see a direct way to obtain these new identities from first principles, a clue may be found in the Ward identity discussion of \cite{Moore:2010bu}, we leave this for future work.

Another interesting type of identities that follow from our calculation relate temperature derivatives of correlation functions, to other correlation functions. This is due to the fact that we can directly derive Kubo formulas for the temperature derivatives of the thermal functions $P'_{i} = - T P_{iT}$ and compare them to the (previously obtained) Kubo formulas for the thermal function differentiated w.r.t the temperature.
For instance we can obtain $P'_1$ from the two point function of $T^{xy}T^{tt}$:
\begin{equation}
P'_1 = - \frac{1}{T}  \lim_{k_x,k_y \rightarrow 0 } \frac{\del^2}{\del k_x \del k_y} G^{xy,tt} (k,-k)\biggr|_{k_0=0}\ .
\end{equation}
This implies the following identity:
\begin{equation}
\boxed{ \lim_{\vec{k} \rightarrow 0 } \left[ \frac{\del^2}{\del k_x \del k_y} G^{xy,tt} (k)
+ \frac{\del^2}{\del k_z^2} G^{xy,xy} (k) \right]=
T \frac{d}{dT}  \left[ \lim_{k_z \rightarrow 0 } \frac{\del^2}{\del k_z^2} G^{xy,xy} (k) \right]}\ .
\end{equation}
Similar identities relate temperature differentiation of two point functions to three point functions (e.g. comparing $P''_1$ from $\langle T^{xx}T^{tt}T^{tt} \rangle$ to the Kubo formulas we already found for $P_1\sim G^{xy,xy}$ differentiated twice w.r.t the temperature).
As we mentioned for the previous type of Kubo formulas, we do not see how to derive these from first principles. We do however hope that this type of Kubo formulas, when studied for charged hydrodynamics, will provide for a more direct study of the relation between the second derivative of the vortical conductivity (equation \eqref{anom:vort_cond},\eqref{anom:vort_kubo}) with respect to the temperature $\del^2_T \xi_\omega \sim C_2 \sim \del^2_T \langle JT \rangle$ and the gravitational anomaly coefficient $\beta \sim \langle TTJ \rangle$ (see \cite{Landsteiner:2011cp,Landsteiner:2011iq,Chapman:2012my} and second footnote in \cite{Banerjee:2012iz} for details on this relation, \cite{Jensen:2012kj} for a proof using hydrodynamics on cones). It is possible that for such a calculation the study of higher order hydrodynamics is required.

\end{document}